\documentclass[aps,prb,footinbib,showpacsshowkeys,amssymb,amsmath,superscriptaddress]{revtex4-1}
\pdfoutput=1
\usepackage{graphicx,graphics}
\usepackage[export]{adjustbox}
\usepackage{verbatim}
\usepackage[font=small]{caption}
\usepackage{braket}
\usepackage{float}
\usepackage{breqn}
\usepackage{subfig}
\usepackage{color}
\usepackage{hyperref}
\usepackage{enumerate}
\makeatletter
\let\cat@comma@active\@empty
\makeatother

\begin{document}
	\title{A Study of Curvature Theory for Different Symmetry Classes of Hamiltonian}
	
	\author{Y R Kartik}
	\thanks{yrkartik@gmail.com}
	\author{Ranjith R Kumar}
			\thanks{ranjith.btd6@gmail.com}
	\author{S Rahul}
			\thanks{rahulastronomer02@gmail.com}
	\affiliation{Theoretical Sciences Division, Poornaprajna Institute of Scientific Research, Bidalur	Bengaluru, 562164, India.}
	\affiliation{Graduate Studies, Manipal Academy of 
	Higher Education, Madhava Nagar, Manipal-576104, India.}
	\author{Sujit Sarkar}
		 \thanks{sujit.tifr@gmail.com}
		\affiliation{Theoretical Sciences Division, Poornaprajna Institute of Scientific Research, Bidalur	Bengaluru, 562164, India.}

	\date{\today} 
	
\begin{abstract}
We 
study and present the results of curvature for 
different symmetry classes (BDI, AIII and A) model 
Hamiltonians and also present the transformation of model 
Hamiltonian from one distinct symmetry class to other 
based on the curvature property. We observe the mirror 
symmetric curvature 
for the Hamiltonian with BDI symmetry class but there is 
no evidence of such behavior for Hamiltonians of AIII 
symmetry class. We show the origin of torsion and its 
consequences on the parameter space of topological phase 
of the system. We find the evidence of
torsion for the Hamiltonian 
of A symmetry class. We present Serret-Frenet equations for all model Hamiltonians in $\mathbf{R}^3$ space. To the best of our knowledge, this is the 
first application of curvature theory to the model 
Hamiltonian of different symmetry classes which 
belong to the topological state of matter.
\end{abstract}
	
\maketitle

\section{Introduction}	
Symmetry and topology are two prominent branches of 
physics that reveal many interesting features. It is 
believed that these two branches are always in agreement 
with each other 
\cite{chiu2016classification,hasan2010colloquium,sato2017topological,
altland1997nonstandard,stanescu2016introduction1,senthil2015symmetry,bernevig2013topological}. 
Before the discovery 
of topological phases of matter, Landau theory of 
symmetry breaking was considered as the prominent tool to 
characterize the phases of matter. But thereafter the 
concept got modified. There is no order parameter in 
topological states of system. However, if a system is 
invariant under some symmetry, it gives rise to  
invariant quantities. These invariants can be used to 
characterize the topological states of matter. Based on 
these invariants, a system can be classified into ten 
distinct symmetry classes. Out of these ten 
non-interacting symmetry classes, only a few exhibits 
topological nature in 1D \cite{altland1997nonstandard}. 
Recently there are some interesting studies which 
involves the interplay and relations between different 
symmetry classes \cite{rahul2019interplay1,10.1080,velasco2019classification,sarkar2018quantization}.\\
Differential geometry deals with the study of problems 
by means of differential calculus, integral calculus and 
linear algebraic techniques \cite{toponogov2006differential,bar2010elementary,abbena2017modern}.
 Differential geometry is significant mathematical structure of general theory 
of relativity by using which the concept of manifold, curved 
space-time, gravity can explained much efficiently 
\cite{carroll2004spacetime,schutz1980geometrical,schutz2009first}. 
There are some notable works which explained PT 
symmetric systems through differential geometry  
\cite{zhang2018differential}.\\
Curvature study is an important step in differential 
analysis of the system and it is effectively used in 
 thermodynamics and many-body systems to explain its 
 nature. Curvature is a tool to measure how 
 curved a curve is. In other words, curvature measures 
 the extent to which a curve deviates from a straight 
 line. For a unit speed curve $\gamma(t),$ where t is a 
 parameter, curvature $\kappa(t)$ at a point is defined 
 to be $||\ddot{\gamma}(t)||$ \cite{pressley2010elementary}. 
 The main motivation is to explain the many body system 
 in a more rigorous manner.
Curves and angles are the effective ways of expressing 
the geometric properties of a physical system 
\cite{wilson2007curved,mokrousov2014geometric,montiel2009curves}. 
Torsion is the natural quantity which is 
associated with the curvature. It affects the 
periodicity, spin wave dynamic and structural defects of 
the system \cite{sheka2015torsion,lima2017torsion,de1994spin}. 
Torsion also have a significant role in the dynamics of 
the adiabatic system, transport properties and 
bulk-boundary correspondence in the topological state of 
matter \cite{hughes2013torsional,khaidukov2018chiral}.\\ 
The geometrical studies of condensed matter systems have 
been an interesting area of research which has rapidly 
picked up pace when the principles of topology and 
geometry were involved in the foundations of quantum 
condensed matter systems \cite{kolodrubetz2013classifying,zulkowski2012geometry}. The physics 
of geometry of curves in $R^3$ with spins in connection with the dynamics of
classical Heisenberg ferromagnetic system under different contexts has already 
been explored in
the literature (see for example,  \cite{lakshmanan2011fascinating,lakshmanan1976dynamics,lakshmanan1977continuum}\\
The main motivation of this work is to study a few model 
Hamiltonians which belong to different symmetry 
classes from the perspective of curved space theory of differential geometry \cite{bar2010elementary,abbena2017modern}.
This paper is organized in the following manner. In Sec 
\ref{ModelHam} we introduce the model Hamiltonian and 
present a detailed analysis of symmetry class 
Hamiltonians. In Sec \ref{Sec3} we present the 
characteristics and behavior of parameter space curves 
with a detailed analysis of differential geometric study 
of curvature. Here we try to analyze the origin of 
torsion and its consequences for the present model 
Hamiltonian.
  
\section{ Basic Model Hamiltonian}\label{ModelHam}
Here we consider eight model Hamiltonians belonging to  
different symmetry classes \cite{rahul2019interplay1,10.1080}. Our model Hamiltonian is 
expressed as
\begin{equation}
H=H_0+H_{eff},
\end{equation}
where $H_0$ is the initial Hamiltonian and $H_{eff}$ is 
effective part of the Hamiltonian which is responsible for the transformation from one symmetry class to other.
Here initial Hamiltonian $H_0$ is a 1D non-interacting 
topological insulator.
We can write our Hamiltonian in the BdG format as
\begin{equation}
H_{BdG}(k)=\chi^{(1)}\left(\begin{matrix}
0&& 1\\
1&& 0\\
\end{matrix}\right)+\chi^{(2)}\left(\begin{matrix}
0&& i\\
-i&& 0\\
\end{matrix}\right)+\chi^{(3)}\left(\begin{matrix}
1&& 0\\
0&& -1\\
\end{matrix}
\right).\end{equation}
The components can be written as, $\chi^{(1)}=0,\chi^{(2)}=\Delta\sin k$ and 
$\chi^{(3)}=\mu+2t\cos k$. 
The effective term ($H_{eff}$) is momentum dependent, 
in the following form
 $H_{eff}=\delta_1k\sigma_{x}+\delta_2k\sigma_{y}+\delta_3k\sigma_{z}=\delta_i(\vec{k_i}.\vec{\tau}_i)$, where $k_1=k_2=k_3=k$
 (the detailed study is presented in the reference \cite{rahul2019interplay1}). 
 We consider a very specific type of effective term which 
 is of much theoretical interest.
 The results of this study may motivate researchers in 
 quantum simulation studies to look for this type of 
 effective term and consequences of their effect on the 
 topological state of matter 
 \cite{georgescu2014quantum,sarkar2015quantum,sarkar2015existence}.\\\\\\\\
 
 \textbf{(1). Hamiltonian $H^{(1)}(k)$}(When $\delta_1=\delta_2=\delta_3=0$)\\
 
Here the effective part of the Hamiltonian is zero. So 
the Hamiltonian in Pauli basis can be written as
   	 \begin{equation}  H_{k}^{(1)}=  2\Delta\sin k \sigma_{y}+(2t\cos k+ \mu ) 
   	 \sigma_{z} .\end{equation}  
   	 Presenting the Hamiltonian in matrix from as 
   	 \begin{equation}
   	 \mathcal{H}^{(1)}(k) = \left( \begin{matrix}
   	   2t\cos(k)+\mu && 2i\Delta\sin(k)\\
   	   -2i\Delta\sin(k) && -2t\cos(k)-\mu\\
   	   \end{matrix}\right).
   	   \label{int1}
   	   \end{equation} 

\textbf{(2). Hamiltonian $H^{(2)} (k)$ }(When 
$\delta_1=\delta_3=0,\delta_2\neq0$)\\
Here the effective term is added to the $\sigma_y$ 
component of the Hamiltonian. It can be written in terms 
of Pauli basis as
   \begin{equation}  H_{k}^{(2)}= 
(2\Delta\sin k+\delta_2 k) \sigma_{y}+(2t\cos k+ \mu ) \sigma_{z}.\end{equation}
   Writing the Hamiltonian in the matrix form
   \begin{equation}
   \mathcal{H}^{(2)}(k) =\left(\begin{matrix}
   2t\cos(k)+\mu && 2i\Delta\sin(k)+i\delta_2 k\\
   -2i\Delta\sin(k)-i\delta_2 k &&- 2t\cos(k)-\mu
   \end{matrix} \right).
   \end{equation}

\textbf{(3). Hamiltonian $H^{(3)}(k)$}(When $\delta_3\neq 0,\delta_1=\delta_2=0$)\\
 Here the effective term is added to the $\sigma_x$ 
 component of the Hamiltonian. It can be written in terms 
 of Pauli basis as
\begin{equation}  H_{k}^{(3)}=  
2\Delta\sin k \sigma_{y}+(2t\cos k+ \mu + \delta_3 k) \sigma_{z} .\end{equation}  
Presenting the Hamiltonian in matrix from as 
\begin{equation}
\mathcal{H}^{(3)}(k) = \left( \begin{matrix}
  2t\cos(k)+\mu+\delta_3 k && 2i\Delta\sin(k)\\
  -2i\Delta\sin(k) && -2t\cos(k)-\mu-\delta_3 k\\
  \end{matrix}\right).
  \label{int1}
  \end{equation} 

\textbf{(4). Hamiltonian $H^{(4)}(k)$ }(When 
$\delta_1=0,\delta_2\neq 0,\delta_3\neq0$)\\
 Here effective terms are added to both the $\sigma_x$ 
 and $\sigma_y$ components of the Hamiltonian. It can be 
 written in terms of Pauli basis as
   \begin{equation}  H_{k}^{(4)}= (2\Delta\sin k+\delta_{2} k) \sigma_{y}+(-2t\cos k- \mu + \delta_{3} k) \sigma_{z} 
   .\end{equation}
   The Hamiltonian $H^{(4)}(k)$ written in the matrix form as
   \begin{equation}
   \mathcal{H}^{(4)}(k)=\left(  \begin{matrix}
   2t\cos(k)+\mu+\delta_3 k && 2i\Delta\sin(k)+i\delta_2 k\\
   -2i\Delta\sin(k)-i\delta_2 k &&- 2t\cos(k)-\mu-\delta_3 k \\
   \end{matrix} \right).
   \label{int3}
   \end{equation}
   
   \textbf{(5). Hamiltonian $H^{(5)}(k)$ }(When 
   $\delta_1\neq0,\delta_2=\delta_3=0$)\\
    Here effective term is added to the $\sigma_x$ 
    component of the Hamiltonian. It can be 
    written in terms of Pauli basis as
      \begin{equation}  H_{k}^{(5)}= (\delta_1 k) \sigma_{x}+ (2\Delta\sin k) \sigma_{y}+(2t\cos k+ \mu) \sigma_{z} 
   .\end{equation}
      The Hamiltonian $H^{(5)}(k)$ written in the matrix form as
      \begin{equation}
      \mathcal{H}^{(5)}(k)=\left(  \begin{matrix}
      2t\cos(k)+\mu && 2i\Delta\sin(k)+\delta_1 k\\
      -2i\Delta\sin(k)+\delta_1 k && -2t\cos(k)-\mu \\
      \end{matrix} \right).
      \label{int5}
      \end{equation}      
         \textbf{(6). Hamiltonian $H^{(6)}(k)$ }(When 
         $\delta_1\neq0,\delta_2\neq0, \delta_3=0$)\\
          Here effective terms are added to both the $\sigma_x$ and $\sigma_y$ 
          components of the Hamiltonian. It can be 
          written in terms of Pauli basis as
            \begin{equation}  H_{k}^{(6)}= (\delta_1 k) \sigma_{x}+ (2\Delta\sin k+\delta_{2}k) \sigma_{y}+(2t\cos k+ \mu) \sigma_{z} 
         .\end{equation}
            The Hamiltonian $H^{(6)}(k)$ written in the matrix form as
            \begin{equation}
            \mathcal{H}^{(6)}(k)=\left(  \begin{matrix}
            2t\cos(k)+\mu && 2i\Delta\sin(k)+i\delta_{2}k+\delta_1 k\\
            -2i\Delta\sin(k)-i\delta_{2}k+\delta_1 k && -2t\cos(k)-\mu \\
            \end{matrix} \right).
            \label{int5}
            \end{equation}            
               \textbf{(7). Hamiltonian $H^{(7)}(k)$ }(When 
               $\delta_1\neq0,\delta_2=0,\delta_3\neq0$)\\
                Here effective terms are added to both the $\sigma_x$ and $\sigma_z$ 
                components of the Hamiltonian. It can be 
                written in terms of Pauli basis as
                  \begin{equation}  H_{k}^{(7)}= (\delta_1 k) \sigma_{x}+ (2\Delta\sin k+\delta_{3}k) \sigma_{y}+(2t\cos k+ \mu) \sigma_{z} 
               .\end{equation}
                  The Hamiltonian $H^{(7)}(k)$ written in the matrix form as
                  \begin{equation}
                  \mathcal{H}^{(7)}(k)=\left(  \begin{matrix}
                  2t\cos(k)+\mu+\delta_{3}k && 2i\Delta\sin(k)+\delta_1 k\\
                  -2i\Delta\sin(k)+\delta_1 k && -2t\cos(k)-\mu-\delta_{3}k \\
                  \end{matrix} \right).
                  \label{int5}
                  \end{equation}                  
                     \textbf{(8). Hamiltonian $H^{(8)}(k)$ }(When 
                     $\delta_1\neq0,\delta_2\neq0,\delta_3\neq0$)\\
                      Here effective terms are added to the $\sigma_x,\sigma_y$ and $\sigma_z$  
                      components of the Hamiltonian. It can be 
                      written in terms of Pauli basis as
                        \begin{equation}  H_{k}^{(8)}= (\delta_1 k) \sigma_{x}+ (2\Delta\sin k+\delta_{2}k) \sigma_{y}+(2t\cos k+ \mu+\delta_{3}k) \sigma_{z} 
                     .\end{equation}
                        The Hamiltonian $H^{(8)}(k)$ written in the matrix form as
                        \begin{equation}
                        \mathcal{H}^{(8)}(k)=\left(  \begin{matrix}
                        2t\cos(k)+\mu+\delta_{3}k && 2i\Delta\sin(k)+i\delta_{2}k+\delta_1 k\\
                        -2i\Delta\sin(k)-i\delta_{2}k+\delta_1 k && -2t\cos(k)-\mu-\delta_{3}k \\
                        \end{matrix} \right).
                        \label{int5}
                        \end{equation}
  \noindent The addition of the effective term does not 
   affect the Hermitian property of the system.\\
    Basically the Hamiltonian is in the spinless fermion 
    basis. The effective term is also in spinless basis 
    and is momentum dependent. Therefore we justify the 
    physical relevance of the effective term. \\

\noindent The first Hamiltonian $H_1(k) $ is the Kitaev model Hamiltonian \cite{kitaev2001unpaired} which governs the topological state of quantum matter. The other seven
Hamiltonians (i.e. from $H_2(k)$ to $H_8(k) $) are the variant of Kitaev model Hamiltonian. We
consider these additional Hamiltonians in the spirit of 
theoretical studies only. By these model Hamiltonians we study the topological as well as
geometric properties of
quantum condensed matter system upto some extent. \\  
 		\section{A curvature analysis of curves in planar parameter space}\label{Sec3}
 		  Curvature can be defined as the rate of variation of 
 		  the angle that the tangent line is making at a 
 		  particular point. To call a curve as a regular curve, 
 		  it should have a non vanishing tangent line. 
 		Curve theory basically deals with analyzing the basic 
 		properties of the curves. Basic properties include, the 
 		arc length, winding number with curvature and torsion of 
 		the curves \cite{pressley2010elementary}. Topological 
 		invariant quantities, like winding number, Chern number 
 		depend on the topology of the parameter space, where for 
 		a particular topological configuration space, winding 
 		number acquires a definite value, and change in the 
 		winding number leads to the different topological 
 		configuration of the system \cite{berry1985classical}.\\
 		  The  understanding of the curve concept is simplified 
 		  by using the differential geometry tool called curvature $\kappa$.
 		 
\begin{widetext}
	
	\begin{figure}[H]
 		 		\centering
 		 		\includegraphics[width=12cm,height=5cm]{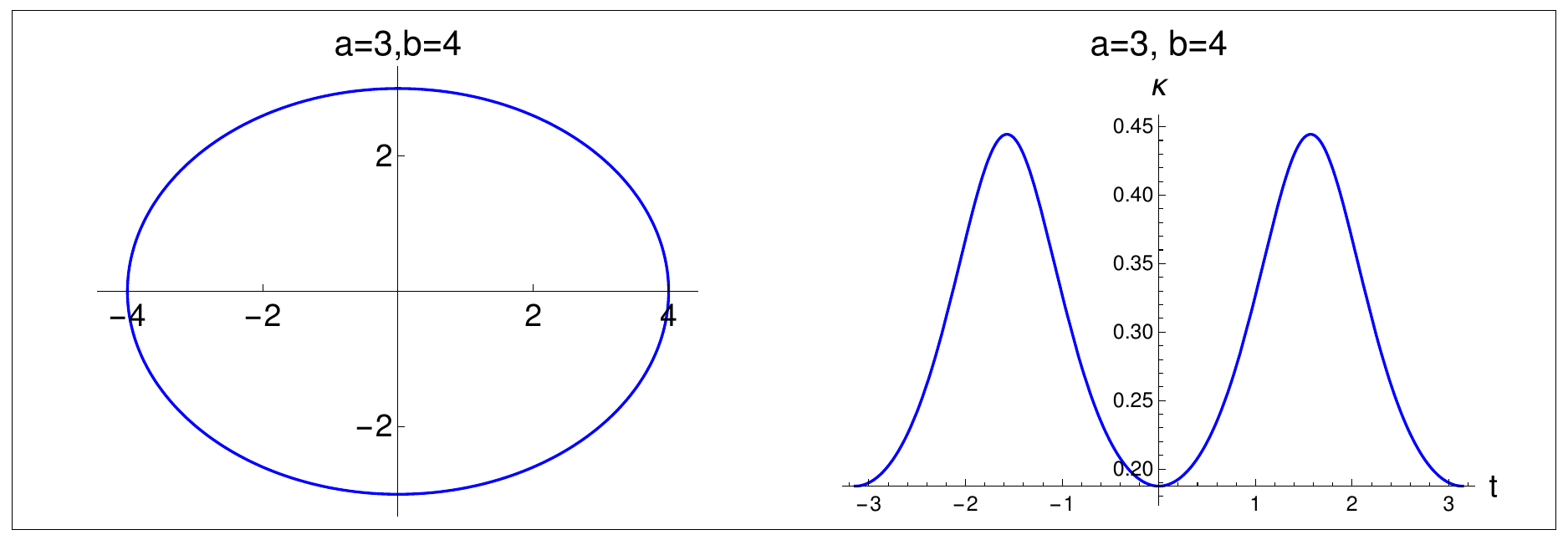}
 		 		\caption{The graphical representation of an ellipse (Left) and the corresponding curvature 
 		plots for the parameter space (Right). Here we observe that the curvature 
 		is minimum at the origin but not vanishing.}
 		 		\label{ellipse}
 		 	\end{figure}
 		 
 		 	\end{widetext}

 		The relation which relates the parameterized curve $c(k)$ 
 		and the curvature $\kappa(t)$ is given by 
 		\cite{ablowitz2003complex} 
 		    \begin{equation}
 		    \kappa(k) = \frac{det(\dot{c(k)},\ddot{c(k)})}{|| \dot{c(k)}||^{3}}\hspace{0.35cm},
 		    \label{kappa}
 		    \end{equation} 
 		    	where dot represents $d/dk$. For a unit speed curve $c:I\longrightarrow \mathbb{R}^2$ 
 		    	where $I=[a,b]$ a closed curve interval. Then 
 		    	$\dot{c}(k)$ gives the velocity vector 
 		    	defined by $(\cos\theta(k), \sin\theta(k))^T$ of 
 		    	an integer multiple of $2\pi$, as the curve is 
 		    	defined in a closed interval. As the angle 
 		    	changes along the curve, the invariant quantity 
 		    	winding number is defined by 
 		    	$\theta(b)-\theta(a)$. If 
 		    	$\theta_1,\theta_2:I\longrightarrow\mathbb{R}$ 
 		    	satisfies the velocity equation. It results as 
 		    	$\theta_1=\theta_2+2n\pi$, where 
 		    	$n\in\mathbb{Z}$.\\
 		    	The velocity term 
 		    	$\dot{c}([a,b])\subset\mathbb{S}_R$, i.e., 
 		    	$\dot{c}(t)>0$ for all $k\in I$ and 
 		    	$\dot{c}(t)=(\dot{c}_1,\dot{c}_2)^T$,\\ 
 		    	$\frac{\dot{c}_{(2)}}{\dot{c}_{(1)}}=\frac{\sin\theta(k)}{\cos\theta(k)}=\tan\theta(k).$ 
 		    	And $\theta(k)=\arctan(\frac{\dot{c}_2(k)}{\dot{c}_1(k)})+2n\pi,n\in\mathbb{Z}$.
 		    	So considering 
 		    	$c:\mathbb{R}\longrightarrow\mathbb{R}^2$  a 
 		    	unit speed vector of a curve with period L and 
 		    	$\theta:\mathbb{R}\longleftarrow\mathbb{R}$ be 
 		    	scalar and winding number is given by \begin{equation}
 		    	w_k=\frac{1}{2\pi}(\theta(L)-\theta(0)).
 		    	\end{equation} where $(\theta(L)-\theta(0))$ is 
 		    	well defined irrespective of the choice if 
 		    	$\theta$. Therefore it is clear from the above 
 		    	equation that to get a complete physical picture 
 		    	of winding number, the study of curve is 
 		    	useful.	
 		 It is well known that the topological system is a closed 
 		 curve which encircles the origin. Geometrically the 
 		 parameter space of a topological system is an ellipse 
 		 and defined as locus of points such that sum of 
 		 distances from the foci is constant.  
 		 	The standard equation of ellipse is given by,  
 		 	$\frac{x^2}{a^2}+\frac{y^2}{b^2}=1,$ where a and b 
 		 	are semi-major and semi-minor axes respectively.
 		 	The parametric equation is given by 
 		$[a(k),b(k)]=(a \cos k,b\sin k)$ where $0\leq k<2\pi $.\\
 		 	The curvature of ellipse is given by 
 		 	 	\cite{bar2010elementary} 
 		\begin{equation}
 		\kappa(k)=\frac{ab}{(b^2\cos^2k+a^2\sin^2k)^{\frac{3}{2}}},\label{kk2}\end{equation} 	
 		 	 	where $a$ and $b$ are semi-major axis and semi-minor 
 		 	 	axis of ellipse respectively(Fig~\ref{ellipse}).
 		 	    From these two parameters we can analyze the 
 		 	    curvature in three different cases.
 		 	
 		 	 	First case: When $a < b$, the curvature is maximum on 
 		 	the semi-major axis ( $-\frac{\pi}{2}$ and 
 		 	$\frac{\pi}{2}$) and it is minimum on the semi-minor 
 		 	axis.\\
 		 	Second case: When $a=b$, the parameter space curve is 
 		 	a circle with the constant curvature.\\
 		 	Third case: When $a>b$, the curvature is minimum on 
 		 	the semi-major axis ($-\frac{\pi}{2}$ and 
 		 	$\frac{\pi}{2}$) and it is maximum on the semi-minor 
 		 	axis \cite{abbena2017modern}.\\
 For a plane unit speed curve $c:I\rightarrow\mathbf{R}^2$, where n(k) and $\kappa(k)$ gives the normal unit vector and curvature of the curve. Then,
\begin{equation}
(\dot{\nu(k)},\dot{n(k)})=(\nu(k),n(k))\left(  \begin{matrix}
      0 && -\kappa(k)\\
      \kappa(k)&&0\\
      \end{matrix} \right)
\end{equation}
defines the relation $\nu=k.n$ and $\dot{n}=-\kappa.\nu$, where $\nu$ is equal to $\dot{c(k)}$ \cite{bar2010elementary}. This is called Frenet equation and gives the information about the curvature properties of the curve $c(k).$\\\\
For a non-vanishing curve $c(k)$with a non-vanishing curvature $\kappa(k)$, torsion is given by \cite{pressley2010elementary}
\begin{equation}
 \tau=\frac{(\dot{c}(k)\times\ddot{c}(k))\cdot\dddot{c}(k)}{||\dot{c}(k)\times\ddot{c}(k)||^2}.
\end{equation}
 Our model Hamiltonian is 
 written in the Pauli spin basis. Naturally the quantity 
 torsion gives the curl of the derivatives of the curve. 
 This results in the in the curve opening and helical 
 motion on the addition of the effective term $\alpha k$. \\
To understand the kinematic properties of curve $c:\mathbf{R}\rightarrow\mathbf{R}^3$, we study Serret-Frenet equation for the curve \cite{pressley2010elementary}. For a unit-speed curve $c(k)$ in $\mathbf{R}^3$ curvature explains the failure of a curve to be a straight line and torsion explains the failure of a line to be a planar. Serret-Frenet formula describe the derivative of tangent $(T)$, normal $(N)$ and binormal $(B)$ unit vectors with respect to arc-length of the parameter of the curve ($s$)\cite{pressley2010elementary}. i.e.,
\begin{eqnarray}
\frac{d\mathbf{T}}{ds}&=&\kappa \mathbf{N}\nonumber\\
\frac{d\mathbf{N}}{ds}&=&-\kappa \mathbf{T}+\tau \mathbf{B}\nonumber\\
\frac{d\mathbf{B}}{ds}&=&-\tau \mathbf{N}\label{fs1}
\end{eqnarray}
Here $\dot{\mathbf{B}}$ is perpendicular to $\mathbf{T}$. Being perpendicular to both $\mathbf{T}$ and $\mathbf{B}$, $\dot{\mathbf{B}}$	must be parallel to $\mathbf{N}$. It is to be noted that, torsion ($\tau$) exists only for a curve with non-zero curvature.
Eq.~\ref{fs1} is known as Serret-Frenet equation and gives the better understanding of the geometric properties of the system. One can also write the matrix representation of the Serret-Frenet equation as follows\cite{pressley2010elementary}.
\begin{equation}
\frac{d}{ds}(X)=\left(  \begin{matrix}
      0 && \kappa(k)&&0\\
      -\kappa(k)&&0&&\tau(k) \\
      0&&-\tau(k)&&0
      \end{matrix} \right)(X),
\end{equation}
where $X=(T,N,B)^{T}$. By expressing $\frac{dT}{ds},\frac{dN}{ds}$ and $\frac{dB}{ds}$ in terms of T, N and B one can get skew-symmetric matrix and it follows
that the vectors T, N and B are orthonormal for all values of arc-length parameter (s).
 \section{Different Symmetry classes and their nature}
Different symmetry classes have already been studied and discussed in the literature extensively \cite{rahul2019interplay1,10.1080,velasco2019classification}. Here, in Table~\ref{es} we discuss it very briefly which are directly involve with the present study. 
 \begin{widetext}
 \begin{table}[H]
 \begin{center}
 \begin{tabular}{ |c|c|c|c| } 
 \hline
 Symmetry&Relation& Operator& Nature \\ 
 \hline
 \hline
 &&&\\
 Time reversal & $[\mathcal{T},H]=0$& $\mathcal{T}=\mathcal{K}$& Reverses the arrow of time\\
 ($\mathcal{T}$) &  $\mathcal{T}HT^{-1}=H$&$\mathcal{T}^2=1$ & $\mathcal{T} : t\longrightarrow-t$\\
 \hline
 &&&\\		 
 Particle-hole &  $\{\mathcal{C},H\}=0$&$\mathcal{C}=\sigma_{x}\mathcal{K}$& Transformation between electron and holes\\
 ($\mathcal{C}$)&&&\\
  &  $\mathcal{C}H\mathcal{C}^{-1}=-H$&$\mathcal{C}^2=1$ & (within certain energy range)\\
 \hline
 &&&\\
Chiral  &  $\{\mathcal{S},H\}=H$&$\mathcal{S}=\sigma_{x}$ & Symmetric spectrum of the Hamiltonian\\
($\mathcal{S}$) &&&\\		
  &$\mathcal{S}H\mathcal{S}^{-1}=-H$& $\mathcal{S}^2=1$ & \\		 		 
 \hline
 \end{tabular} 
 \end{center}
 \caption{Properties of symmetry operators which are related with the present study.}
 	\label{es}
 \end{table}
 \end{widetext}
\noindent\textbf{Time-reversal symmetry (TR):}
Time-reversal symmetry is the transformation which is anti-unitary in nature. The time-reversal operator just reverses the sign of momentum but does not affect the position. 
It is equivalent to the complex conjugate operator ($\mathcal{K}$).
\begin{equation}
\mathcal{T}x\mathcal{T}^{-1}=x , \;\;\; \mathcal{T}k\mathcal{T}^{-1}=-k,\;\;\; 
\mathcal{T}i\mathcal{T}^{-1}=-i .
\end{equation}
Time reversal operator is the product of unitary
($U$) and complex conjugate operators, i.e. $\mathcal{T}=U\mathcal{K}$.
The square of the time-reversal operator equals negative of identity which yields to Kramer's degeneracy.
According to that one state is time-reversal of another and every state is doubly degenerate.
Thus the system becomes time-reversal invariant 
\cite{kotetestopological,shiozaki2014topology,stanescu2016introduction1}.\\
\textbf{Particle-hole (PH) symmetry:}
The particle-hole operator is an anti-unitary operator and
with the presence of this symmetry each Eigen-function $\Psi$ with $E>0$ has its particle-hole 
reversed partner, $\mathcal{C}\Psi$ with $E<0$. The PH symmetry is the intrinsic property of mean field theory of superconductivity.\\
\textbf{Chiral symmetry:}
Chiral symmetry $(S)$ or sub-lattice symmetry is the product of time-reversal operator $(T)$ and particle-hole operator $(C)$ . 
Based on the behavior
of Hamiltonian with the TR, PH and chiral symmetries, it is classified into 10 symmetry classes. \\
In Table~\ref{AZ}, we present the different symmetry classes to characterize the topological 
states of the system for different dimension ($d$). The first column present the different symmetry
classes, the second, third and fourth column present are respectively for the time-reversal, particle-hole
and charge conjugation symmetry. The rest of the table is for the dimensionality ($d$) and the
topological index system.($ Z$ and $Z_2 $). Here we mention very briefly the topological characterization
of the system, the detail discussion can be found in the following references \cite{rahul2019interplay1,10.1080,velasco2019classification}.\\  
\noindent Topological states of matter are characterized by the 
 presence of time reversal, chiral and charge conjugation 
 symmetries. They are 
  classified into different symmetry classes based on these  symmetry operators. The edge 
 state in the topological systems are protected by the 
 time reversal symmetry ($\mathbb{T}: t\longrightarrow -t$) 
 and time reversal symmetry (commutes with the 
 Hamiltonian. i.e.,  $[\mathbb{H},\mathbb{T}]=0$), 
 chiral symmetry (i.e., chiral operator anti-commutes with 
 Hamiltonian, $\{\mathbb{S},\mathbb{H}\}=0$) and particle-hole operator (anti-commutes with the Hamiltonian  $\{\mathbb{S},\mathbb{H}\}=0$) decides 
 whether the system is topological or not. i.e.,  The 
 table \ref{AZ} presents the condition and classification 
 of different symmetry classes.  We observe that our model Hamiltonians 
 belong to three different (BDI,
 AIII and A) symmetry classes. We present our results of different symmetry classes in the next section. 
 \begin{widetext}
 \begin{table}[H]
 			\centering
 			\includegraphics[width=12cm,height=6cm]{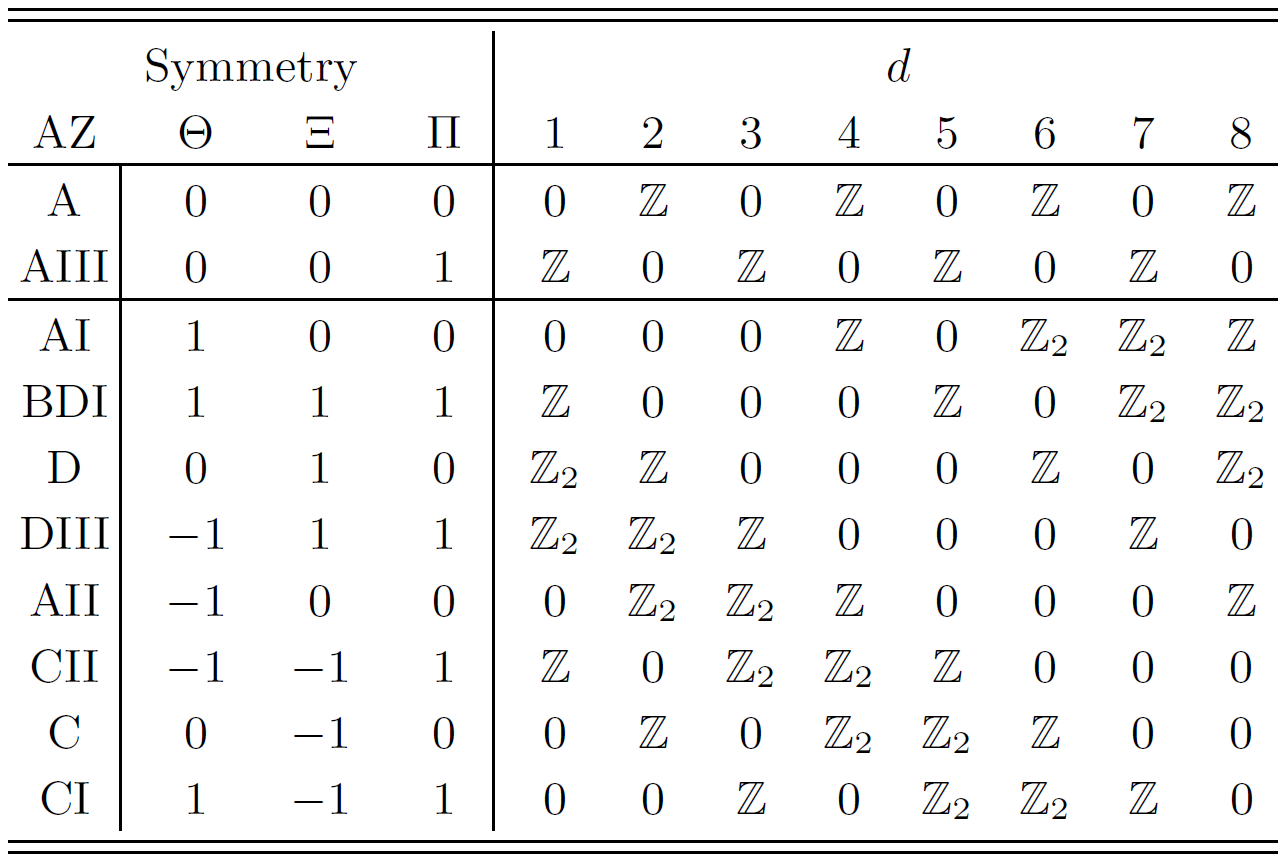}
 			\caption{Ten fold symmetry class for a topological system. 
Here $\Theta$ is the time reversal, $\Xi$ is the particle-hole, $\Pi$ is the charge 
conjugation symmetry and d is the dimensionality of system respectively.}
 			\label{AZ}
 		\end{table}
 		 \end{widetext}
\subsection*{\textbf{Results of BDI symmetry class}} BDI symmetry 
 class is characterized by the commutation of time 
 reversal ($\mathbf{T}$) operator with the Hamiltonian 
 anti-commutation of other two operators like 
 particle-hole ($\mathbb{C}$) and chiral($\mathbb{S}$) 
 with the Hamiltonian (Eq. \ref{AZ}). Here the Hamiltonians 
 $H^{(1)}(k)$ and $H^{(2)}(k)$ belongs to the BDI class 
 \cite{rahul2019interplay1}. The Hamiltonian $H^{(1)}(k)$ 
 is topological in nature. The Hamiltonian $H^{(2)}(k)$ 
 shows the topologically trivial behavior. Now we study the curvature properties of these Hamiltonians.\\\\
	\textbf{(1) $H^{(1)}(k)$ Hamiltonian}\\
	Here we present the results of differential geometric 
	study based on curve theory for the BDI Hamiltonians.
	The matrix form of the model Hamiltonian is 
	\begin{equation}
	\mathcal{H}^{(1)} (k)= \left( \begin{matrix}
	2t\cos(k)+ \mu && 2i\Delta\sin(k)\\
	-2i\Delta\sin(k) && -2t\cos(k)- \mu \\
	\end{matrix}\right). \label{kitaev11}
	\end{equation}
	 Here the set of possible parametric equations are
	 \begin{eqnarray}
\chi^{(1)}(H^{(1)}(k))&=&0\nonumber\\
\chi^{(2)}(H^{(1)}(k))&=&2\Delta\sin k,\nonumber\\
\chi^{(3)}(H^{(1)}(k))&=&2t\cos k+\mu, 
	 \end{eqnarray}
	
	   	    $H_{BdG}$ Hamiltonian in the pseudo spin basis is 
 
\begin{equation}
	   	   H^{(1)}(k)=\chi^{(2)}(H^{(1)}(k))\sigma_{y}+\chi^{(3)}(H^{(1)}(k)) \sigma_{z}.
\end{equation}
In terms of vectors, one can write the above equation as $H_{BdG}=\bar{\chi}(k).\bar{\tau}$, where $\bar{\tau}$ are the Pauli spin matrices acting in the particle-hole (Nambu) basis of $H_{BdG}$ \cite{niu2012majorana}.
The energy dispersion relation is, 
$E^{(1)}(k)=\sqrt{(2t\cos k+\mu)^2+(2\Delta\sin k)^2}$.\\

\begin{widetext}
	\begin{figure}[H]
			\centering
			\includegraphics[width=12cm,height=5cm]{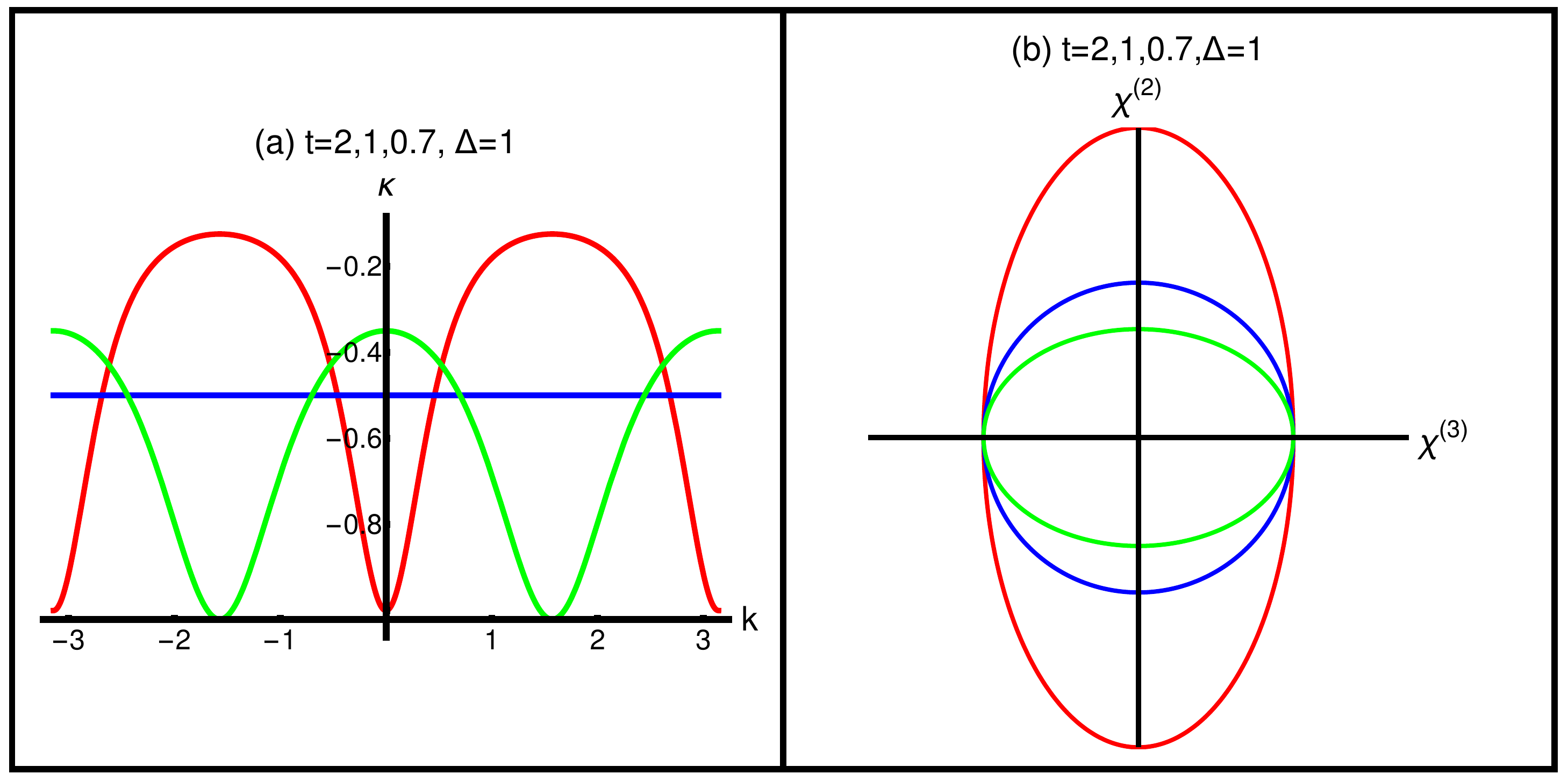}
			\caption{The left figure represents the plots 
			of curvature with k for the values 
			$\gamma$=2, 1, 0.7 for red, blue and green 
			respectively. The right figure represents 
			corresponding parameter plots for the value 
			$\mu=0$}
			\label{kk1}
		\end{figure}
	
	\end{widetext}

	Considering the parametric equation of the 
	Hamiltonian $H^{(1)} (k)$ in the matrix form 
	\begin{equation}
	c(k)=\left[  \begin{matrix}
	2\Delta\sin k\\
		2t\cos k+\mu
	\end{matrix}\right],
	 \dot{c}(k)=\left[  \begin{matrix}
	2\Delta\cos k\\
	-2t\sin k
	\end{matrix}\right], \ddot{c}(k)=\left[  \begin{matrix}
		-2\Delta\sin k\\
	-2t\cos k

	\end{matrix}\right].	\end{equation}\label{eq34}
	Curvature is given by
	\begin{eqnarray}
	\kappa=\frac{det[\dot{c},\ddot{c}]}{||\dot{c}||^3} &=&\frac{det\left(\begin{matrix}
		2\Delta\cos k&&-2\Delta\sin k \\
		-2t\sin k && -2t\cos k
		\end{matrix}\right)}{(\sqrt{4 t^2 \sin^2k+4\Delta^2\cos^2k})^3}\nonumber\\
&=&\frac{-2t\Delta}{(\sqrt{t^2 \sin^2k+\Delta^2\cos^2k})^3}
	\label{k1} \hspace{0.35cm}.
	\end{eqnarray}
		Fig. \ref{kk1} represents the curvature plot for 
		Hamiltonian $ H^{(1)}(k)$. 
			The parameter space curve for the Hamiltonian 
			$ H^{(1)}(k)$ is nothing but an ellipse 
			(fig.\ref{ellipse}) due to the mathematical 
			structure of the parametric equation.
	
	\noindent For the value $\mu=0$, the system remains in the 
	topological state. We can study the curvature of 
	parameter space curve for all Hamiltonians. We cannot 
	characterize the topological and non-topological 
	states of the Hamiltonian from the curvature study. 
	The reason for this is, the curvature expression does 
	not include the term $\mu$. From the above general 
	discussion on the ellipse we can characterize the 
	parameter space curve of the $H^{(1)}(k)$ Hamiltonian 
	into similar three cases which is described bellow. 
	This is completely a theoretical study to understand 
	the behavior of the parameter space curve of the 
	model Hamiltonians from the perspective of 
	differential geometry.\\	First case: When 
	$t<\Delta$, the curvature is maximum on the 
	semi-major axis ( $-\frac{\pi}{2}$ and 
	$\frac{\pi}{2}$) and it is minimum on the semi-minor 
	axis.\\
			Second case: When $t=\Delta$, the 
			parameter space curve is a circle with the 
			constant curvature.\\
			Third case: When $t>\Delta$, the 
			curvature is minimum on the semi-major axis 
			($-\frac{\pi}{2}$ and $\frac{\pi}{2}$) and it 
			is maximum on the semi-minor axis.\\\\
 		\noindent	\textbf{(2) $H^{(2)}(k)$ 
 		Hamiltonian}\\
	Hamiltonian $H^{(2)} (k)$ can be written in the 
	matrix form as
	\begin{equation}
		\mathcal{H}^{(2)}(k) =\left(\begin{matrix}
			2t\cos(k)+\mu && 2i\Delta\sin(k)+i\delta_2 k\\
			-2i\Delta\sin(k)-i\delta_2 k && -2t\cos(k)-\mu
		\end{matrix} \right).
	\end{equation}
	Here the set of parametric equations are
	  \begin{eqnarray}
\chi^{(1)}(H^{(2)}(k))&=&0\nonumber\\
\chi^{(2)}(H^{(2)}(k))&=&2\Delta\sin k+\delta_2 k,\nonumber\\
	\chi^{(3)}(H^{(2)}(k))&=&2t\cos k+\mu.
	  \end{eqnarray}
	    $H_{BdG}$ Hamiltonian in the pseudo spin basis is\cite{niu2012majorana}
  
\begin{equation}
	   H ^{(2)}(k)=\chi^{(2)}(H^{(2)}(k))\sigma_{y}+\chi^{(3)}(H^{(2)}(k)) \sigma_{z}
	    \end{equation}
	    The energy dispersion relation, $E^{(2)}(k)=\sqrt{(2t\cos k+\mu)^2+(2\Delta\sin k+\delta_2 k)^2}.$\\
	
Hence the curvature for $H^{(2)}(k)$ is
	\begin{eqnarray}
		\kappa&=&\frac{det\left[\begin{matrix}
							2\Delta\cos k+\delta_2 && -2\Delta\sin k\\
				-2t\sin k && -2t\cos k

			\end{matrix}\right]}{(\sqrt{(2t \sin k)^2+(2\Delta\cos k+\delta_2)^2})^3}\nonumber\\
&=&\frac{-4t\Delta-
2\delta_2 t\cos k}{(\sqrt{(2t \sin k)^2+(2\Delta\cos k+\delta_{2})^2})^3}\hspace{0.35cm}. \label{k2}
	\end{eqnarray}
Eq. \ref{k2} is the analytical expression of the curvature for the Hamiltonian $H^{(2)} (k)$. 
\begin{widetext}
\begin{figure}[H]
	\centering
	\includegraphics[width=12cm,height=8cm]{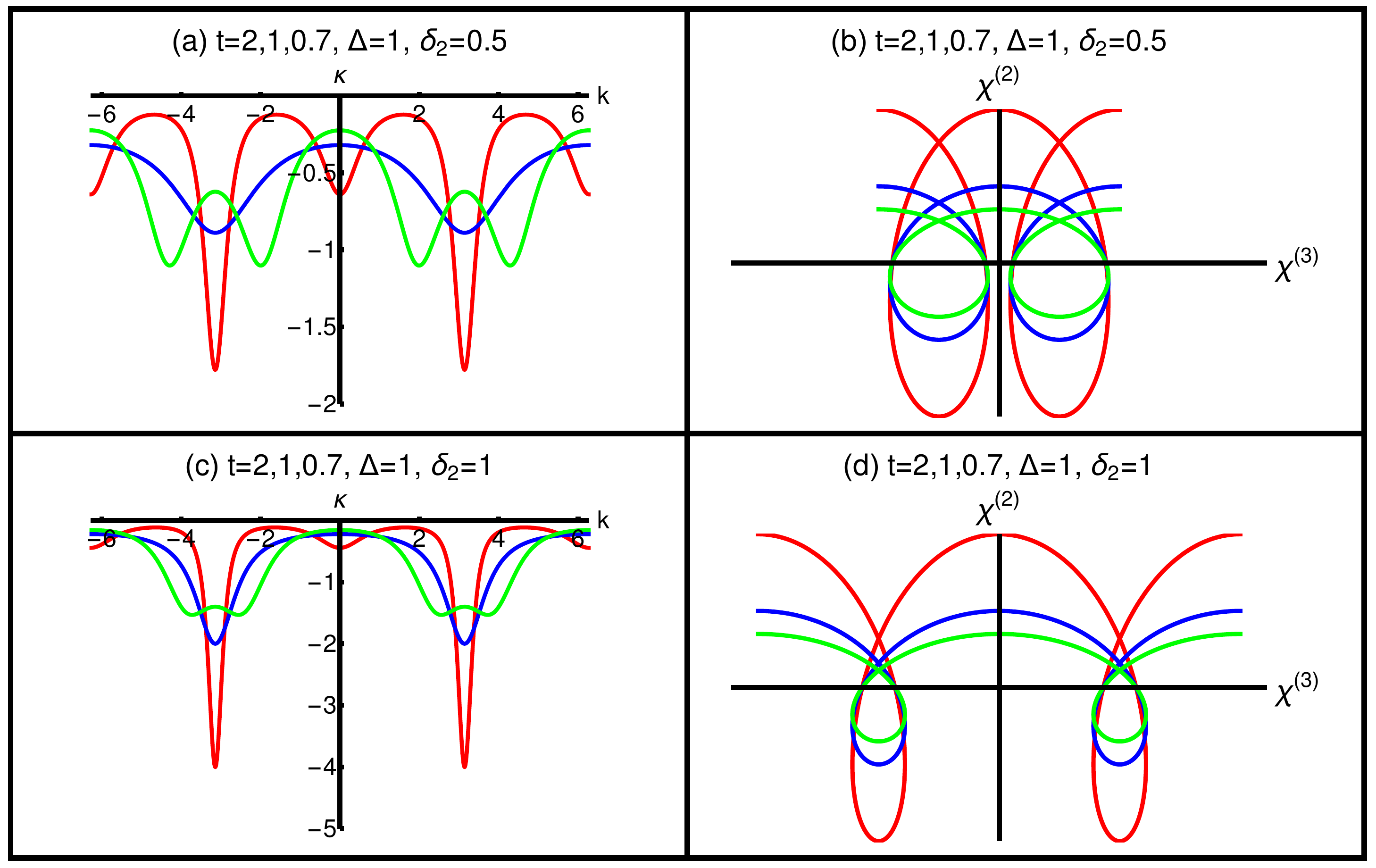}
	\caption{a) Plot of curvature ($\kappa$) with k for 
	$\delta_{2}=0.5$. b) Corresponding parameter plots 
	for the curvature plot a. c) Plots of curvature 
	($\kappa$) with k for $\delta_{2}=1$. d) 
	Corresponding parameter plots for the curvature plot 
	c. In all plots the red, blue and green colors 
	represents $t$=2,1,0.7 respectively }
	\label{H2}
\end{figure}
\end{widetext}
Fig \ref{H2} consists of of two panels for two different values of $\delta_{2}$. The upper and lower panel represents the parameter space $\delta_{2}=0.5$ and $\delta_{2}=1$ respectively.
Each panel consists of two figures, the left one is for curvature and the right one is for corresponding parameter space curves. We observe that with the increasing value of $\delta_{2}$, the curvature also increases. Here we observe an interesting feature that 
the curvature as well as parameter plots are mirror 
symmetric about $\kappa$ axis. This is true for 
both Hamiltonians of BDI symmetry class. The parameter 
space curve splits into two as we increase the value of 
$\delta_{2}$.\\
The parameter space curves of Hamiltonian $H^{(2)}(k)$ 
resembles the cycloidal pattern due to its mathematical 
structure. 
The general expression of the cycloid is given by 
\cite{abbena2017modern} 
\begin{equation}
    Cyc[a,b](t) = (at-b\sin t , a-b \cos t).\label{cc}
\end{equation}
    In general the cycloid is classified into two 
    categories depending on the values of coefficients. 
    Suppose in Eq.\ref{cc}, if $a<b$, then the cycloid is 
    prolate and if $a>b$, it is curate.
    From this classification, we can assign our 
    Hamiltonian $H^{(2)}(k)$, as prolate since the 
    prolate cycloid is self-interacting and also it 
    satisfies the condition $a<b$.\\

\noindent One can notice that the presence of effective 
term changes the properties of differential geometry 
which we study the curvature properties in the parameter 
space. Based on the strength of the effective term, the 
parameter space curve behaves as simple curve with non 
closed, self intersecting conditions.\\
For this BDI symmetry class, we have presented curvature 
study of two different Hamiltonians. Hamiltonian 
$H^{(1)}(k)$ is the model Hamiltonian without effective 
term. In  Hamiltonian $H^{(2)}(k)$, the effective term is 
added to the $\sigma_{y}$ component. Here, in both the 
cases, the curvature is mirror symmetric about the 
$\kappa$ axis.\\\\
\subsection*{\textbf{Results of AIII symmetry class}}
AIII symmetry is characterized by the absence of time 
reversal and particle-hole symmetry. But it obeys chiral 
symmetry condition (Fig \ref{AZ}).  
AIII symmetry class contains two Hamiltonians 
$H^{(3)}(k)$ and $H^{(4)}(k)$. Both Hamiltonians are 
topologically trivial in one dimension and satisfies all the symmetry 
properties.\\\\
	\textbf{(3) $H^{(3)}(k)$ Hamiltonian}\\
	The matrix form of the Hamiltonian $H^{(3)}(k)$ is
	\begin{equation}
		\mathcal{H}^{(3)}(k) = \left( \begin{matrix}
			2t\cos(k)+\mu+\delta_3 k && 2i\Delta\sin(k)\\
			-2i\Delta\sin(k) && -2t\cos(k)-\mu-\delta_3 k\\
		\end{matrix}\right).
		\label{eq5}
	\end{equation} 
 Here the set of possible parametric equations are
 \begin{eqnarray}
\chi^{(1)}(H^{(3)}(k))&=&0\nonumber\\
\chi^{(2)}(H^{(3)}(k))&=&2\Delta\sin k\nonumber\\
\chi^{(3)}(H^{(3)}(k))&=&2t\cos k+\mu+\delta_3 k, 
 \end{eqnarray}
   $H_{BdG}$ Hamiltonian in the pseudo spin basis is \cite{niu2012majorana}

\begin{equation}
  H^{(3)}(k)= \chi^{(2)}(H^{(3)}(k)) \sigma_{y}+\chi^{(3)}(H^{(3)}(k)) \sigma_{z}.
   \end{equation}
   The energy dispersion relation is 
   $E^{(3)}(k)=\sqrt{(2t\cos k+\mu+\delta_2 k)^2+(2\Delta\sin k)^2}$.\\

	The Curvature of the Hamiltonian $H^{(3)}(k)$ is 
	\begin{eqnarray}
	\kappa(k)&=&\frac{det\left(\begin{matrix}
				2\Delta\cos k && -2\Delta\sin k\\
		-2t\sin k+\delta_3 && -2t\cos k
\end{matrix}\right)}{(\sqrt{(-2t \sin k+\delta_3)^2+4\Delta^2\cos^2k})^3}\nonumber\\
&=&\frac{-4t\Delta+2\delta_3\Delta\sin k}{(\sqrt{(2t \sin k+\delta_3)^2+(2 \alpha\cos k)^2})^3}\hspace{0.35cm}.
	\end{eqnarray}
\begin{widetext}
Fig \ref{H1} consists of of two panels for two different values of $\delta_{3}$. The upper and lower panel represents the the parameter space $\delta_{3}=0.5$ and $\delta_{3}=1$ respectively.
Each panel consists of two figures, the left one is for curvature and the right one is for corresponding parameter space curves. We observe that with the increasing value of $\delta_{3}$, the curvature also increases.
\begin{figure}[H]
	\centering
	\includegraphics[width=12cm,height=8cm]{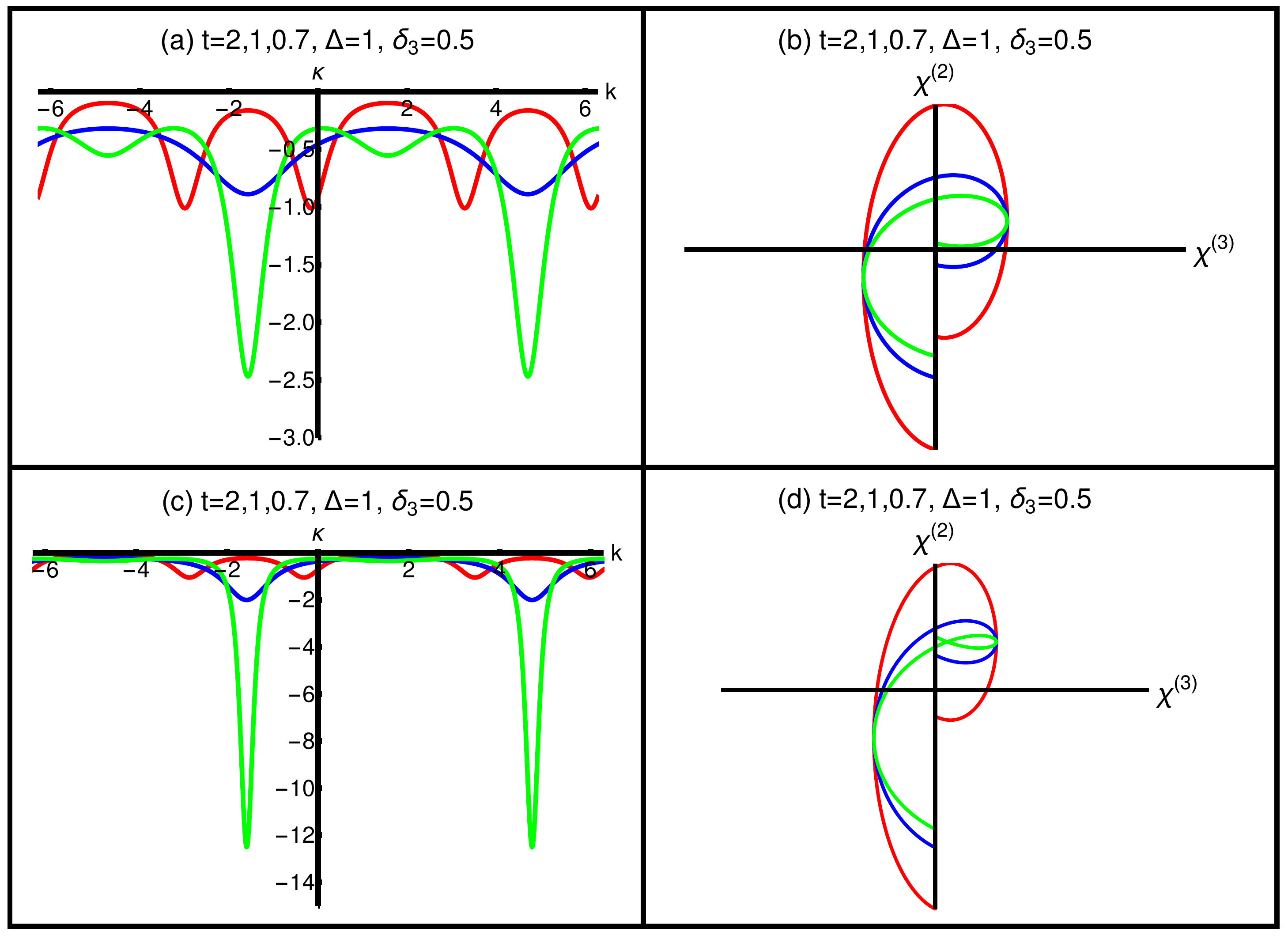}
	\caption{a) Plots of curvature ($\kappa$) with k for 
	$\delta_{2}=0.5$. b) Corresponding parameter plots 
	for the curvature plot a. c) Plots of curvature 
	($\kappa$) with k for $\delta_{2}=1$. d) 
	Corresponding parameter plots for the curvature plot 
	c. In all plots the red, blue and green colors 
	represents $t$=2,1,0.7 respectively}
	\label{H1}
\end{figure} 

\end{widetext}
It reveals in this study that the AIII symmetry class 
lacks the mirror symmetry about $\kappa$ axis. As 
the value of $\delta_{1}$ increases, the peaks become 
steep but their position is unaltered. 
As the previous case, the curvature expression is 
independent of the term $\mu$. The increase in the 
strength of the effective term results in decrease of 
curvature near $k=0$.\\ 
	For the Hamiltonian $H^{(3)} (k)$, the parameter 
	space curve is also a prolate cycloid because it is 
	open  self-intersecting.\\
	 From the curvature studies for this parameter space 
	 curve of Hamiltonian $H^{(3)} (k)$, it reveals that 
	 the curvature at the points ($-\pi$ and $\pi$) on 
	 the semi-major axis is maximum and the curvature on 
	 the semi-minor axis is minimum. When the effective 
	 term changes its sign, the parameter space curves as 
	 well as curvature plots forms mirror symmetric image 
	 \cite{rahul2019interplay1}.\\  
	 
	\textbf{(4) $H^{(4)}(k)$ Hamiltonian.}\\
	Hamiltonian $H^{(4)} (k)$ can be written in the matrix form as
	\begin{equation}
	\mathcal{H}^{(4)}(k)=\left(  \begin{matrix}
	2t\cos(k)+\mu+\delta_3 k && 2i\Delta\sin(k)+i\delta_2 k\\
	-2i\Delta\sin(k)-i\delta_2 k && -2t\cos(k)-\mu-\delta_3 k \\
	\end{matrix} \right)
	\label{h3}
	\end{equation}
	    Here the set of possible parametric equations are
	    \begin{eqnarray}
	   \chi^{(1)}(H^{(4)}(k))&=&0\nonumber\\
	   \chi^{(2)}(H^{(4)}(k))&=&2\Delta\sin k+\delta_2k,\nonumber\\
	   \chi^{(3)}(H^{(4)}(k))&=&2t\cos k+\mu+\delta_3 k.
	    \end{eqnarray}
	     $H_{BdG}$ Hamiltonian in the pseudo spin basis is \cite{niu2012majorana}
	      \begin{equation}
	    H(k)^{(4)}=\chi^{(2)}(H^{(4)}(k)) \sigma_{y}+\chi^{(3)}(H^{(4)}(k)) \sigma_{z}
	    \end{equation}
 
	     The energy dispersion relation, 
	     $E^{(4)}(k)=\sqrt{(2\Delta\sin k+\delta_2 k)^2+(2t\cos k+\mu+\delta_3 k)^2}.$\\
	Curvature is given by
	\begin{eqnarray}
	\kappa(k)&=&\frac{\textrm{Det}\left[\begin{matrix}
				2\Delta\cos k+\delta_2 && -2\Delta\sin k\\
		-2t\sin k+\delta_3 && -2t\cos k
\end{matrix}\right]}{(\sqrt{(-2t \sin k+\delta_3)^2+(2\Delta\cos k+\delta_2)^2})^3}\nonumber\\
&=&\frac{-4t\Delta-2(\delta_3\Delta\sin k+\delta_2 t\cos k)}{(\sqrt{(-2t \sin k+\delta_3)^2+(2\Delta\cos k+\delta_2)^2})^3}.
	\label{cyc6}
	\end{eqnarray}
	Eq.\ref{cyc6} is an analytic expression of the 
	curvature for the Hamiltonian $H^{(4)}(k)$.
	\begin{widetext}
	\begin{figure}[H]
		\centering
		\includegraphics[width=12cm,height=8cm]{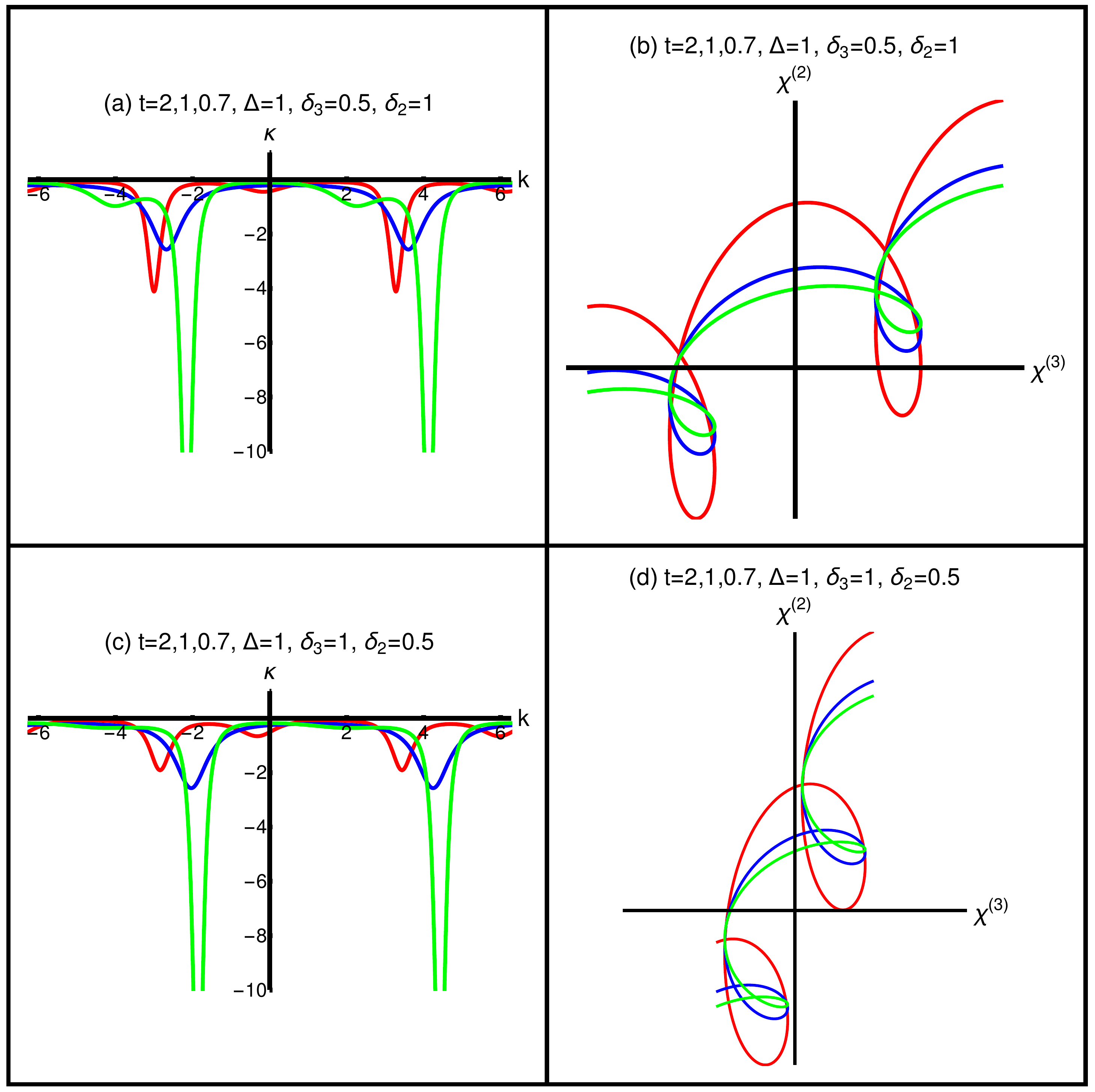}
		\caption{a) Plots of curvature ($\kappa$) with k 
		for $\delta_{1}=0.5$. b) Corresponding parameter 
		plots for the curvature plot a. c) Plots of 
		curvature ($\kappa$) with k for $\delta_{2}=0.5$. 
		d) Corresponding parameter plots for the 
		curvature plot c. In all plots the red, blue and 
		green colors represents $t$=2,1,0.7 
		respectively}
		\label{H3}  
	\end{figure}
	\end{widetext}
Fig \ref{H3} consists of of two panels for two different values of $\delta_{2}$ and $\delta_{3}$. The upper and lower panel represents the the parameter space $\delta_{2}=1,\delta_{3}=0.5$ and $\delta_{2}=0.5,\delta_{3}=1$ respectively.
Each panel consists of two figures, the left one is for curvature and the right one is for corresponding parameter space curves. We observe that with the increasing value of $\delta_{2}$, the curvature also increases.
	It clearly shows the evidence of divergence 
	in the curvature plots. $H^{(4)}(k)$ shows the 
	asymmetry nature same as $H^{(3)}(k)$ Hamiltonian. 
	For $H^{(4)}(k)$ Hamiltonian, parameter space curves 
	forms cycloidal pattern but in a very arbitrary way. 
	There is no specific way of orientation. The 
	corresponding curvature shows the non-topological 
	state. Based on the strength of $\delta_2$ and 
	$\delta_3$ there arises divergence characters at the 
	BZ boundary values. 
	The curvature plots shows the divergence at BZ 
	boundary regions i.e.,$-\pi$ and $\pi$. \\
	In AIII symmetry class, we have presented two model 
	Hamiltonians. Hamiltonian $H^{(3)}(k)$ is contains 
	the effective term in the $\sigma_{z}$ part and 
	Hamiltonian $H^{(4)}(k)$ contains effective term both 
	in $\sigma_{y}$ and $\sigma_{z}$ components. Both of 
	these Hamiltonians show distorted curves where 
	curvature lacks mirror symmetry about the $\kappa$ 
	axis.\\
	Both BDI as well as AIII symmetry classes have 
	distinct geometric properties. Through curvature 
	study we can analyze the nature of parameter space, 
	cycloidal motion of the parameter space with and without the 
	addition of effective term. When the effective term is added to the $\sigma_{y}$ or $\sigma_{z}$ component of the Hamiltonian, the system remains in the $\mathbf{R^2}$ space and we observe only curvature. But the cycloidal motion of the $\mathbf{R^2}$ parameter space is nothing other than the helical motion in the $\mathbf{R^3}$ space. Hence we consider the $\mathbf{R^3}$ space to investigate the torsional effect of effective term on the model Hamiltonian.
 
\subsection*{\textbf{Results of $\mathbf{A}$ symmetry class}} Symmetry 
 class A is characterized by the absence of time 
 reversal ($\mathbf{T}$), 
 particle-hole ($\mathbb{C}$) and chiral($\mathbb{S}$) 
 with the Hamiltonian  \ref{AZ}. Here the Hamiltonians 
 $H^{(5)}(k),H^{(6)}(k),H^{(7)}(k)$ and $H^{(8)}(k)$ belong to the A class 
 \cite{rahul2019interplay1}. These Hamiltonians 
 shows the topologically trivial behavior for a one-dimensional system.\\\\
	\textbf{(5) $H^{(5)}(k)$ Hamiltonian.}\\
Hamiltonian $H^{(5)} (k)$ can be written in the matrix form as
\begin{equation}
\mathcal{H}^{(5)}(k)=\left(  \begin{matrix}
2t\cos(k)+\mu && 2i\Delta\sin(k)+\delta_1 k\\
-2i\Delta\sin(k)+\delta_1 k && 2t\cos(k)+\mu \\
\end{matrix} \right).
\label{h3}
\end{equation}
Here the set of possible parametric equations are
\begin{eqnarray}
\chi^{(1)}(H^{(5)}(k))&=&\delta_{1}k,\nonumber\\
\chi^{(2)}(H^{(5)}(k))&=&2\Delta\sin k,\nonumber\\
\chi^{(3)}(H^{(5)}(k))&=&2t\cos k+\mu.
\end{eqnarray}
$H_{BdG}$ Hamiltonian in the pseudo spin basis is \cite{niu2012majorana} 
\begin{equation}
H(k)^{(5)}=\chi^{(1)}(H^{(5)}(k)) \sigma_{x}+\chi^{(2)}(H^{(5)}(k)) \sigma_{y}+\chi^{(3)}(H^{(5)}(k)) \sigma_{z}.
\end{equation}
The energy dispersion relation, 
$E^{(5)}(k)=\sqrt{(\delta_1 k)^2+(2\Delta\sin k)^2+(-2t\cos k-\mu)^2}.$\\
The parameter space of $H^{(5)}(k)$ belongs to $\mathbf{R}^3$ space and forms the circular helix as
\begin{equation}
helix\left[ a,b\right] (k)=(a\cos(k),a\sin(k),bk)
\end{equation}
where $a$ is the radius and $b$ is the slope of the helix (here for all cases we take $\Delta=t$ to achieve unit speed curve properties). The projection of $\mathbf{R}^3$ onto $\mathbf{R}^2$ maps the helix onto a circle.\\
Here the curve is 
\begin{equation}
	c(k)=\left[  \begin{matrix}
	\delta_1 k\\
	2\Delta\sin k\\
		2t\cos k+\mu
	\end{matrix}\right],
		 \dot{c}(k)=\left[  \begin{matrix}
	\delta_1\\
		2\Delta\cos k\\
		-2t\sin k
		\end{matrix}\right], \ddot{c}(k)=\left[  \begin{matrix}
		0\\
			-2\Delta\sin k\\
		-2t\cos k	
		\end{matrix}\right].
	\end{equation}
	And thus the curvature $\kappa=||\ddot{c}(k)||=2$ which represents the non vanishing curvature. Hence it is possible to find normal vector for all values of $k$. Thus
	\begin{equation}
n(k)=\frac{\ddot{c}(k)}{\kappa(k)}=\frac{1}{2}\left[  \begin{matrix}
		0\\
			-2\Delta\sin k\\
		-2t\cos k	
		\end{matrix}\right].
	\end{equation}
	Binormal vector is given by
	\begin{eqnarray}
b(k)=\dot{c}\times n(k)&=&\left[  \begin{matrix}
\delta_1\\
		2\Delta\cos k\\
		-2t\sin k
		\end{matrix}\right]\times\frac{1}{2}\left[  \begin{matrix}
		0\\
			-2\Delta\sin k\\
		-2t\cos k	
		\end{matrix}\right]\nonumber\\&=&\frac{1}{2}\left[  \begin{matrix}
				-4t\Delta\\
					-2\Delta\delta_1\sin k\\
				-2t\delta_1\cos k			
				\end{matrix}\right].
	\end{eqnarray}
		\begin{figure}[H]
						\centering
						\includegraphics[width=12cm,height=12cm]{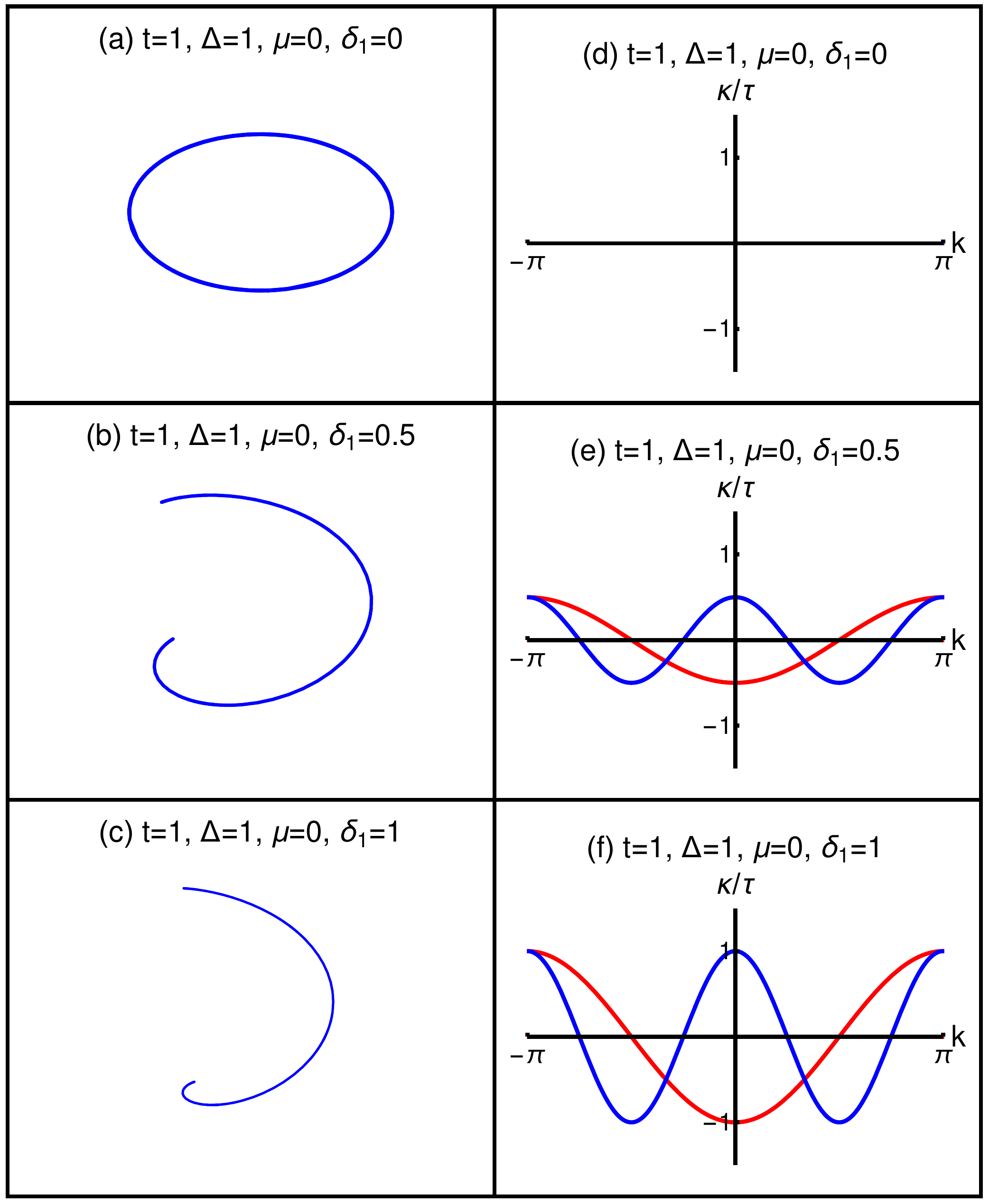}
						\caption{(Left) Parameter 
						plots for the Hamiltonian $H^{(5)}(k)$. (Right) Plots of 
						curvature ($\kappa$) and torsion ($\tau$) with k for $t=\Delta=1$ and $\alpha=0,0.5$ and 1. The red and blue lines in the right panel represent the corresponding normal curvatures as well as torsion respectively.}
						\label{H5}  
					\end{figure}	
	The torsion is given by
	\begin{eqnarray}
\left\langle \dot{n}(k),b(k)\right\rangle=\left\langle \frac{1}{2}\left[  \begin{matrix}
		0\\
			-2\Delta\cos k\\
		2t\sin k	
		\end{matrix}\right],\frac{1}{2}\left[  \begin{matrix}
				-4t\Delta\\
					-2\Delta\delta_1\sin k\\
				-2t\delta_1\cos k			
				\end{matrix}\right]\right\rangle 
				=t\Delta\delta_1.
	\end{eqnarray}
		Thus the curvature as well as the torsion gives constant values for the $H^{(5)}(k)$.\\
			By using Eq.~\ref{fs1}, Serret-Frenet equations can be written to $H^{(5)}(k)$ Hamiltonian as,
			\begin{eqnarray}
		\dot{T}(k)&=&\left[  \begin{matrix}
				0\\
					-2\Delta\sin k\\
				-2t\cos k	
				\end{matrix}\right],\nonumber\\
		\dot{N}(k)&=&-2\left[  \begin{matrix}
			\delta_1\\
				2\Delta\cos k\\
				-2t\sin k
				\end{matrix}\right]+\frac{t\Delta\delta_1}{2}\left[  \begin{matrix}
								-4t\Delta\\
									-2\Delta\delta_1\sin k\\
								-2t\delta_1\cos k
								\end{matrix}\right], \nonumber\\
		\dot{B}(k)&=&-\frac{t\Delta\delta_1}{2}\left[  \begin{matrix}
				0\\
					-2\Delta\sin k\\
				-2t\cos k	
				\end{matrix}\right].
			\end{eqnarray}
		Thus Serret-Frenet equations to $H^{(5)}(k)$ Hamiltonian gives the understanding about the dynamics of $H^{(5)}(k)$ Hamiltonian. When the $H^{(5)}(k)$ Hamiltonian is projected from $\mathbf{R^3}\rightarrow\mathbf{R^2}$ and one can obtain the $H^{(1)}(k)$ Hamiltonian.\\
		Fig. \ref{H5} represents the study of curvature as well torsion to $H^{(5)}(k)$ Hamiltonian. The left panel indicates the parameter space and the right panel indicates corresponding curvature and torsion for different values of $\delta_{1}$. From the plot it is clear that, with the increasing values of $\delta_{1}$, the amplitude of curvature and torsion also increases. Hence the curvature and torsion are directly proportional to $\delta_{1}$.\\\\			
	\textbf{(6) $H^{(6)}(k)$ Hamiltonian.}\\
Hamiltonian $H^{(6)} (k)$ can be written in the matrix form as
\begin{equation}
\mathcal{H}^{(6)}(k)=\left(  \begin{matrix}
2t\cos(k)+\mu && 2i\Delta\sin(k)+i\delta_{2}k+\delta_1 k\\
-2i\Delta\sin(k)-i\delta_{2}k+\delta_1 k && 2t\cos(k)+\mu \\
\end{matrix} \right).
\end{equation}
Here the set of possible parametric equations are
\begin{eqnarray}
\chi^{(1)}(H^{(6)}(k))&=&\delta_{1}k,\nonumber\\
\chi^{(2)}(H^{(6)}(k))&=&2\Delta\sin k+\delta_{2}k,\nonumber\\
\chi^{(3)}(H^{(6)}(k))&=&2t\cos k+\mu.
\end{eqnarray}
$H_{BdG}$ Hamiltonian in the pseudo spin basis is \cite{niu2012majorana}
 \begin{equation}
H(k)^{(6)}=\chi^{(1)}(H^{(6)}(k)) \sigma_{x}+\chi^{(2)}(H^{(6)}(k)) \sigma_{y}+\chi^{(3)}(H^{(6)}(k)) \sigma_{z}.
\end{equation}
The energy dispersion relation, 
$E^{(6)}(k)=\sqrt{(\delta_1 k)^2+(2\Delta\sin k+\delta_{2}k)^2+(-2t\cos k-\mu)^2}.$\\
Here the curve is 
\begin{equation}
	c(k)=\left[  \begin{matrix}
	\delta_1 k\\
	2\Delta\sin k+\delta_{2}k\\
		2t\cos k+\mu
	\end{matrix}\right],
		 \dot{c}(k)=\left[  \begin{matrix}
	\delta_1\\
		2\Delta\cos k+\delta_{2}\\
		-2t\sin k
		\end{matrix}\right], \ddot{c}(k)=\left[  \begin{matrix}
		0\\
			-2\Delta\sin k\\
		-2t\cos k	
		\end{matrix}\right].
	\end{equation}
	And thus the curvature $\kappa=||\ddot{c}(k)||=2$ which represents the non vanishing curvature. Hence it is possible to find normal vector for all values of $k$. Thus
	\begin{equation}
n(k)=\frac{\ddot{c}(k)}{\kappa(k)}=\frac{1}{2}\left[  \begin{matrix}
		0\\
			-2\Delta\sin k\\
		-2t\cos k	
		\end{matrix}\right].
	\end{equation}
	Binormal vector is given by
	\begin{eqnarray}
b(k)=\dot{c}\times n(k)&=&\left[  \begin{matrix}
\delta_1\\
		2\Delta\cos k+\delta_{2}\\
		-2t\sin k
		\end{matrix}\right]\times\frac{1}{2}\left[  \begin{matrix}
		0\\
			-2\Delta\sin k\\
		-2t\cos k	
		\end{matrix}\right]\nonumber\\&=&\frac{1}{2}\left[  \begin{matrix}
				-4t\Delta-2t\delta_{2}\cos k\\
					-2\Delta\delta_1\sin k\\
				-2t\delta_1\cos k			
				\end{matrix}\right].
	\end{eqnarray}
	The torsion is given by
	\begin{eqnarray}
\left\langle \dot{n}(k),b(k)\right\rangle=\left\langle \frac{1}{2}\left[  \begin{matrix}
		0\\
			-2\Delta\cos k\\
		2t\sin k	
		\end{matrix}\right],\frac{1}{2}\left[  \begin{matrix}
				-4t\Delta-2t\delta_{2}\cos k\\
					-2\Delta\delta_1\sin k\\
				-2t\delta_1\cos k			
				\end{matrix}\right]\right\rangle 
				=t\Delta\delta_1.
	\end{eqnarray}
		Thus the curvature as well as the torsion gives constant values for the $H^{(6)}(k)$.
	By using Eq.~\ref{fs1}, Serret-Frenet equations can be written to $H^{(6)}(k)$ Hamiltonian as,
	\begin{eqnarray}
\dot{T}(k)&=&\left[  \begin{matrix}
		0\\
			-2\Delta\sin k\\
		-2t\cos k	
		\end{matrix}\right],\nonumber\\
\dot{N}(k)&=&-2\left[  \begin{matrix}
	\delta_1\\
		2\Delta\cos k+\delta_{2}\\
		-2t\sin k
		\end{matrix}\right]+\frac{t\Delta\delta_1}{2}\left[  \begin{matrix}
						-4t\Delta-2t\delta_{2}\cos k\\
							-2\Delta\delta_1\sin k\\
						-2t\delta_1\cos k
						\end{matrix}\right], \nonumber\\
\dot{B}(k)&=&-\frac{t\Delta\delta_1}{2}\left[  \begin{matrix}
		0\\
			-2\Delta\sin k\\
		-2t\cos k	
		\end{matrix}\right].
	\end{eqnarray}   
	Thus Serret-Frenet equations to $H^{(6)}(k)$ Hamiltonian gives the understanding about the dynamics of $H^{(6)}(k)$ Hamiltonian. When the $H^{(6)}(k)$ Hamiltonian is projected from $\mathbf{R^3}\rightarrow\mathbf{R^2}$ and one can obtain the $H^{(2)}(k)$ Hamiltonian.\\\\
	\textbf{(7) $H^{(7)}(k)$ Hamiltonian.}\\
Hamiltonian $H^{(7)} (k)$ can be written in the matrix form as
\begin{equation}
\mathcal{H}^{(7)}(k)=\left(  \begin{matrix}
2t\cos(k)+\mu+\delta_{3}k && 2i\Delta\sin(k)+\delta_1 k\\
-2i\Delta\sin(k)+\delta_1 k && -2t\cos(k)-\mu-\delta_{3}k \\
\end{matrix} \right),
\end{equation}
Here the set of possible parametric equations are
\begin{eqnarray}
\chi^{(1)}(H^{(7)}(k))&=&\delta_{1}k,\nonumber\\
\chi^{(2)}(H^{(7)}(k))&=&2\Delta\sin k,\nonumber\\
\chi^{(3)}(H^{(7)}(k))&=&2t\cos k+\mu+\delta_{3}k.
\end{eqnarray}
$H_{BdG}$ Hamiltonian in the pseudo spin basis is \cite{niu2012majorana}
\begin{equation}
H(k)^{(7)}=\chi^{(1)}(H^{(7)}(k)) \sigma_{x}+\chi^{(2)}(H^{(7)}(k)) \sigma_{y}+\chi^{(3)}(H^{(7)}(k)) \sigma_{z}.
\end{equation}
The energy dispersion relation, 
$E^{(7)}(k)=\sqrt{(\delta_1 k)^2+(2\Delta\sin k)^2+(2t\cos k+\mu+\delta_{3}k)^2}.$\\
Here the curve is 
\begin{equation}
	c(k)=\left[  \begin{matrix}
	\delta_1 k\\
	2\Delta\sin k\\
		2t\cos k+\mu+\delta_{3}k
	\end{matrix}\right],
		 \dot{c}(k)=\left[  \begin{matrix}
	\delta_1\\
		2\Delta\cos k\\
		-2t\sin k+\delta_{3}
		\end{matrix}\right], \ddot{c}(k)=\left[  \begin{matrix}
		0\\
			-2\Delta\sin k\\
		-2t\cos k	
		\end{matrix}\right].
	\end{equation}
	And thus the curvature $\kappa=||\ddot{c}(k)||=2$ which represents the non vanishing curvature. Hence it is possible to find normal vector for all values of $k$. Thus
	\begin{equation}
n(k)=\frac{\ddot{c}(k)}{\kappa(k)}=\frac{1}{2}\left[  \begin{matrix}
		0\\
			-2\Delta\sin k\\
		-2t\cos k	
		\end{matrix}\right].
	\end{equation}
	Binormal vector is given by
	\begin{eqnarray}
b(k)=\dot{c}\times n(k)&=&\left[  \begin{matrix}
\delta_1\\
		2\Delta\cos k\\
		-2t\sin k+\delta_{3}
		\end{matrix}\right]\times\frac{1}{2}\left[  \begin{matrix}
		0\\
			-2\Delta\sin k\\
		-2t\cos k	
		\end{matrix}\right]\nonumber\\&=&\frac{1}{2}\left[  \begin{matrix}
				-4t\Delta-2\Delta\delta_{3}\sin k\\
					-2\Delta\delta_1\sin k\\
				-2t\delta_1\cos k			
				\end{matrix}\right].
	\end{eqnarray}
	The torsion is given by
	\begin{eqnarray}
\left\langle \dot{n}(k),b(k)\right\rangle=\left\langle \frac{1}{2}\left[  \begin{matrix}
		0\\
			-2\Delta\cos k\\
		2t\sin k	
		\end{matrix}\right],\frac{1}{2}\left[  \begin{matrix}
				-4t\Delta-2\Delta\delta_{3}\sin k\\
					-2\Delta\delta_1\sin k\\
				-2t\delta_1\cos k			
				\end{matrix}\right]\right\rangle 
				=t\Delta\delta_1.
	\end{eqnarray}
		Thus the curvature as well as the torsion gives constant values for the $H^{(7)}(k)$.
	By using Eq.~\ref{fs1}, Serret-Frenet equations can be written to $H^{(7)}(k)$ Hamiltonian as,
	\begin{eqnarray}
\dot{T}(k)&=&\left[  \begin{matrix}
		0\\
			-2\Delta\sin k\\
		-2t\cos k	
		\end{matrix}\right],\nonumber\\
\dot{N}(k)&=&-2\left[  \begin{matrix}
	\delta_1\\
		2\Delta\cos k\\
		-2t\sin k+\delta_{3}
		\end{matrix}\right]+\frac{t\Delta\delta_1}{2}\left[  \begin{matrix}
						-4t\Delta-2\Delta\delta_{3}\sin k\\
							-2\Delta\delta_1\sin k\\
						-2t\delta_1\cos k
						\end{matrix}\right], \nonumber\\
\dot{B}(k)&=&-\frac{t\Delta\delta_1}{2}\left[  \begin{matrix}
		0\\
			-2\Delta\sin k\\
		-2t\cos k	
		\end{matrix}\right].
	\end{eqnarray}
	Thus Serret-Frenet equations to $H^{(7)}(k)$ Hamiltonian gives the understanding about the dynamics of $H^{(7)}(k)$ Hamiltonian. When the $H^{(7)}(k)$ Hamiltonian is projected from $\mathbf{R^3}\rightarrow\mathbf{R^2}$ and one can obtain the $H^{(3)}(k)$ Hamiltonian.\\\\
	\textbf{(8) $H^{(8)}(k)$ Hamiltonian.}\\
Hamiltonian $H^{(8)} (k)$ can be written in the matrix form as
\begin{equation}
\mathcal{H}^{(8)}(k)=\left(  \begin{matrix}
2t\cos(k)+\mu+\delta_{3}k && 2i\Delta\sin(k)+i\delta_{2}k+\delta_1 k\\
-2i\Delta\sin(k)-i\delta_{2}k+\delta_1 k && 2t\cos(k)+\mu+\delta_{3}k \\
\end{matrix} \right).
\end{equation}
Here the set of possible parametric equations are
\begin{eqnarray}
\chi^{(1)}(H^{(8)}(k))&=&\delta_{1}k,\nonumber\\
\chi^{(2)}(H^{(8)}(k))&=&2\Delta\sin k+\delta_{2}k,\nonumber\\
\chi^{(3)}(H^{(8)}(k))&=&2t\cos k+\mu+\delta_{3}k.
\end{eqnarray}
$H_{BdG}$ Hamiltonian in the pseudo spin basis is \cite{niu2012majorana} 
\begin{equation}
H(k)^{(8)}=\chi^{(1)}(H^{(8)}(k)) \sigma_{x}+\chi^{(2)}(H^{(8)}(k)) \sigma_{y}+\chi^{(3)}(H^{(8)}(k)) \sigma_{z}.
\end{equation}
The energy dispersion relation, 
$E^{(8)}(k)=\sqrt{(\delta_1 k)^2+(2\Delta\sin k+\delta_{2}k)^2+(2t\cos k+\mu+\delta_{3}k)^2}.$\\
Here the curve is 
\begin{equation}
	c(k)=\left[  \begin{matrix}
	\delta_1 k\\
	2\Delta\sin k+\delta_{2}k\\
		2t\cos k+\mu+\delta_{3}k
	\end{matrix}\right],
		 \dot{c}(k)=\left[  \begin{matrix}
	\delta_1\\
		2\Delta\cos k+\delta_{2}\\
		-2t\sin k+\delta_{3}
		\end{matrix}\right], \ddot{c}(k)=\left[  \begin{matrix}
		0\\
			-2\Delta\sin k\\
		-2t\cos k	
		\end{matrix}\right].
	\end{equation}
	And thus the curvature $\kappa=||\ddot{c}(k)||=2$ which represents the non vanishing curvature. Hence it is possible to find normal vector for all values of $k$. Thus
	\begin{equation}
n(k)=\frac{\ddot{c}(k)}{\kappa(k)}=\frac{1}{2}\left[  \begin{matrix}
		0\\
			-2\Delta\sin k\\
		-2t\cos k	
		\end{matrix}\right].
	\end{equation}
	Binormal vector is given by
	\begin{eqnarray}
b(k)=\dot{c}\times n(k)&=&\left[  \begin{matrix}
\delta_1\\
		2\Delta\cos k+\delta_{2}\\
		-2t\sin k+\delta_{3}
		\end{matrix}\right]\times\frac{1}{2}\left[  \begin{matrix}
		0\\
			-2\Delta\sin k\\
		-2t\cos k	
		\end{matrix}\right]\nonumber\\&=&\frac{1}{2}\left[  \begin{matrix}
				-4t\Delta-2t\delta_{2}\cos k-2\Delta\delta_{3}\sin k\\
					-2\Delta\delta_1\sin k\\
				-2t\delta_1\cos k			
				\end{matrix}\right].
	\end{eqnarray}
	The torsion is given by
	\begin{eqnarray}
\left\langle \dot{n}(k),b(k)\right\rangle&=&\left\langle \frac{1}{2}\left[  \begin{matrix}
		0\\
			-2\Delta\cos k\\
		2t\sin k	
		\end{matrix}\right],\frac{1}{2}\left[  \begin{matrix}
				-4t\Delta-2t\delta_{2}\cos k-2\Delta\delta_{3}\sin k\\
					-2\Delta\delta_1\sin k\\
				-2t\delta_1\cos k			
				\end{matrix}\right]\right\rangle\nonumber\\ 
				&=&t\Delta\delta_1.
	\end{eqnarray}
		Thus the curvature as well as the torsion gives constant values for the $H^{(8)}(k)$.
	By using Eq.~\ref{fs1}, Serret-Frenet equations can be written to $H^{(8)}(k)$ Hamiltonian as,
	\begin{eqnarray}
\dot{T}(k)&=&\left[  \begin{matrix}
		0\\
			-2\Delta\sin k\\
		-2t\cos k	
		\end{matrix}\right],\nonumber\\
\dot{N}(k)&=&-2\left[  \begin{matrix}
	\delta_1\\
		2\Delta\cos k+\delta_{2}\\
		-2t\sin k+\delta_{3}
		\end{matrix}\right]+\frac{t\Delta\delta_1}{2}\left[  \begin{matrix}
						-4t\Delta-2t\delta_{2}\cos k-2\Delta\delta_{3}\sin k\\
							-2\Delta\delta_1\sin k\\
						-2t\delta_1\cos k
						\end{matrix}\right], \nonumber\\
\dot{B}(k)&=&-\frac{t\Delta\delta_1}{2}\left[  \begin{matrix}
		0\\
			-2\Delta\sin k\\
		-2t\cos k	
		\end{matrix}\right].
	\end{eqnarray}
Thus Serret-Frenet equations to $H^{(8)}(k)$ Hamiltonian gives the understanding about the dynamics of $H^{(8)}(k)$ Hamiltonian. When the $H^{(8)}(k)$ Hamiltonian is projected from $\mathbf{R^3}\rightarrow\mathbf{R^2}$ and one can obtain the $H^{(4)}(k)$ Hamiltonian.\\
Thus it is very clear that the projection of $\mathbf{R^3}\rightarrow\mathbf{R^2}$ space ($\chi_2-\chi_3$ parameter space) signals the changes in the geometrical properties of the model Hamiltonian. In the $\mathbf{R^3}$ space the Hamiltonian belongs to symmetry class A, but when it projected to $\mathbf{R^2}$ space,it belongs to either BDI or AIII symmetry class. It is very important to notice that, under the given conditions, the Hamiltonians of symmetry , class A, show same curvature and torsion. But the Hamiltonians belong to BDI and AIII symmetry class have different curvature expressions.
\subsection*{Geodesic properties of the curve for $H^{(5)}(k)$ Hamiltonian}
Geodesics are the shortest path between two points in a surface. Geodesics always have a constant speed. Sometimes geodesics can be expressed as geodesic curvature ($k_g$). Hence as a part of curvature study, we consider a unit-speed curve on a circular cylinder which actually forms a helix on surface. It is interesting that the intersection of a cylinder the plane perpendicular to its rulings is always a geodesic. Here we consider $H^{(5)}(k)$ Hamiltonian and calculate the geodesic by geometrical operations.\\
Local isometry is the quantity which can give a clear understanding about this. For a unit cylinder $W$ with the conditions $x^2+y^2=1$, there always exists geodesic with the circles obtained by intersecting $W$ with planes parallel to x-y plane. Because of the local isometric property, one can connect the points $(u,v,0)$ of the x-y plane to the points $(cos u, sin u, v)$  of the $W$ plane. This makes a geodesic from x-y plane to the geodesic on $W$. The line which is not parallel to the y-axis in the x-y plane gives the equation $y=mx+c$,  where $m$ and $c$ are constants. Parameterizing the line by $x=k$ and $y=mk+c$ we get $c(k)=(cos (k), sin (k), mk+c)$ which is nothing other than the similar helix considered in $H^{(5)}(k)$. Here we clearly give the geodesic curve for $H^{(5)}(k)$.\\
Let there be a circular cylinder,
\begin{equation}
W=\{X=(\chi_1,\chi_2,\chi_3)\in R^3|\chi_2^2+\chi_3^2=1,\chi_1=k,k\in R\}.
\end{equation}
Here we consider $H^{(5)}(k)$ Hamiltonian with the condition $\mu=0,t=\Delta=1/2$ and $\delta_{1}=1$. The minimum condition for a curve $c:I\rightarrow W$ on $W$ to be a geodesic is that the curve $c(k)$ should be inclined on $W$. Let the curve $c(k)$ be a geodesic on the circular cylinder $W$.\\
Now $\dot{c}(k)=\frac{dc}{dk}=V_1$. If the angle between $V_1$ and $\frac{d}{d\chi_3}$ is $\phi(k)$, then for every $k$ \cite{geodesic1},
\begin{equation}
\langle V_1,\frac{d}{d\chi_1}=\cos(\phi(k))\rangle
\end{equation}
By taking covariant derivatives with respect to $V_1$
\begin{equation}
\langle D_{V_1}V_1,\frac{d}{d\chi_1}\rangle+\langle V_1D_{V_1},\frac{d}{d\chi_1}\rangle=-\sin(\phi(k))\frac{d\phi}{dk}
\end{equation}
or in other words
\begin{equation}
\langle k_1V_2,\frac{d}{d\chi_1}\rangle=-\sin(\phi(k))\frac{d\phi}{dk}
\end{equation}
where $V_2=\frac{\ddot{c}(k)}{||\ddot{c}(k)||},||\ddot{c}(k)||=k_1$. Then $\langle\ddot{c},\frac{d}{d\chi_1}\rangle=-\sin(\phi(k))\frac{d\phi}{dk}$. Here the curve $c(k)$ is a unit speed curve (under given conditions) and a geodesic on the circular helical, hence we get $\ddot{c}(k)=\lambda N$. For a $N$ vector area defined by $N_p=(p_1,p_2,p_3......p_{n-1}=0)$ for $p=(p_1,p_2.....p_n)\in W$ is the unit normal vector area of $W$. So,
\begin{equation}
\langle N,\frac{\partial_n}{\partial k_n}\rangle=\sin(\phi(k))\frac{d\phi}{dk}=0
\end{equation}
Now $\sin(\phi(k))=0$ or $\frac{d\phi(k)}{dk}=0$. So $\frac{d\phi(k)}{dk}=0\Longrightarrow\phi(k)=0$ or $\phi(k)=constant$. This show that the curve is an inclined curve with $\frac{d}{d\chi_1}$ as axis on the circular cylinder $W$.\\
In other way,
\begin{equation}
\langle V_1,\frac{dV_1}{d\chi_1}\rangle=\cos(\phi(k)),\phi(k)\neq\pi/2(\phi=constant)
\end{equation}
Hence, $\langle k_1V_2,\frac{d}{d\chi_1}\rangle=0$. The covariant derivative with respect to $V_1$ is
\begin{equation}
\langle\frac{dV_1}{dk},\frac{d}{d\chi_1}\rangle=0\Longrightarrow\langle N,\frac{d}{d\chi_1}\rangle=0
\end{equation}
It shows $\langle\ddot{c}(k),\dot{c}(k)\rangle=0$ and $N,\dot{c}(k)=0$, where
\begin{eqnarray}
N&=&\lambda\frac{d}{d\chi_1}\wedge\dot{c}(k)\nonumber\\
\ddot{c}&=&\beta\frac{d}{d\chi_1}\wedge\dot{c}(k)
\end{eqnarray}
Then $\ddot{c}(k)=\beta N$, which clearly shows the inclined curve is a geodesic under given parameter space.\\
Here we consider just $H^{(5)}(k)$ Hamiltonian under some particular parameter space to calculate the geodesics. We choose the parameter space in such a way that the curve $c(k)$ remains unit-speed. In other Hamiltonians, it is not possible to achieve unit-speed curve. And we consider unit cylinder with condition $x^2+y^2=1$. This case is only possible in $H^{(5)}(k)$ Hamiltonian. When the effective term is added to either $\sigma_{y}$ or $\sigma_{z}$, the curve fails to be a unit speed curve. As this condition is not possible in other Hamiltonians, we only calculate geodesic curvature to $H^{(5)}(k)$ Hamiltonian.
\subsection{Consequences of effective term and its physical interpretation}	
\noindent The differential geometric analysis of the parameter 
space gives the understanding of the nature of 
Hamiltonians of different symmetry classes. This 
effort successfully explains the curvature study of the 
parameter space with 
the addition of effective term $\alpha k$ and the 
transition of system from topological to 
topologically trivial phase. Curvature and torsion are the integral part of a geometrical system and one can understand the physical system in a better way by the study. In the study 
of space-time geometry, mass is responsible for curvature 
and spin is responsible for torsion \cite{de1994spin}. In 
the same way, for our present model, the dependence of 
momentum vector $k$ in terms $sine$ and $cosine$ are 
responsible for the curvature and effective term $\alpha k$ 
is responsible for the torsional effects. The cycloidal motion in a $\mathbf{R}^2$ space is a cycloid when it is projected to a $\mathbf{R}^3$ space and a unit speed cycloid in a $\mathbf{R}^3$ space is a unit speed circle when it is projected to a $\mathbf{R}^2$ space. This helps to understand the relation between the geometry as well as physics of a quantum condensed mater system.\\
When the same analogy comes to a lattice model, the 
initial Hamiltonian $H_0(k)$ represents a tight binding 
model and the effective term $\delta_{i} k$ represents 
external interaction term which is linear momentum (in 
some cases it is similar to magnetic field). Because of 
the nature of the effective term it gives rise to torsion 
in the lattice system. So it results in the curve opening 
of parameter space and cycloidal motion. \\
 For the tight binding models, this type of torsion 
 results in dislocations and disclinations \cite{de1994spin}. 
 It is similar to the disorder and 
 defect in the crystal lattices. In our Hamiltonians the 
 periodicity of the Bloch space breaks and the system 
 transforms from topological to non-topological phase. 
 This transformation is the result of torsion. Even 
 though the system transforms from topological to trivial 
 phase, the model remains in the respective  symmetry 
 classes (BDI, AIII and A).\\\\      
\noindent\textbf{Conclusion:}
We have presented entirely new and insightful results of 
curvature analysis for different 
symmetry classes, each system class containing  
different Hamiltonians with different topological 
properties. We have shown explicitly  the merits and 
limitations of curvature study in the presence of 
effective term. We have analyzed behavior of system from 
topological to non-topological state with the addition of 
effective term to the model Hamiltonian. We have shown 
explicitly the presence of mirror symmetry for the curvature study of BDI 
symmetry class but that symmetries are absent for the 
AIII and A symmetry classes. We have introduced the concept of 
torsion in topological state of matter and there by 
explained the transformation of system from topological 
to non-topological state and we observed a transformation of symmetry classes, when there is a projection from $\mathbf{R}^3$ space to $\mathbf{R}^2$ space. We have given the geodesic properties of certain Hamiltonian under given conditions. This work 
provides a new perspective on the curvature 
analysis for the topological state of matter.\\\\
\textbf{Acknowledgments}\\
    SS would like to acknowledge DST (EMR/2017/000898) 
    for the funding and RRI library for the books and 
    journals. YRK would like to thank Admar Mutt Education Foundation for the scholarship. The authors would like to acknowledge Dr. 
    R Srikanth, Dr. B S Ramachandra and Prof. C Sivaram who has read this 
    manuscript critically and gave useful suggestions. 
    This research was supported in part by the 
    International Centre for Theoretical Sciences (ICTS) 
    during a visit for participating in the program - 
    Geometry and Topology for Lecturers (Code: ICTS/gtl2018/06).\\  
\bibliography{CurvatureNotes}
\end{document}